\pdfoutput=1
\documentclass[11pt,preprintnumbers,eqsecnum,nofootinbib]{revtex4-1}
\usepackage{amsmath, latexsym, amssymb, hyperref, graphicx, slashed, color, multirow}
\usepackage[T1]{fontenc}
\definecolor{nicered}{rgb}{.7,.1,.1}
\definecolor{nicegreen}{rgb}{.1,.5,.1}
\definecolor{darkblue}{rgb}{0,0,.5}
\hypersetup{colorlinks, citecolor=nicegreen,linkcolor=nicered, urlcolor=darkblue}

\usepackage{cancel}
 \usepackage[normalem]{ulem}
\usepackage{color}
\usepackage{booktabs,siunitx}
\usepackage{booktabs}
\usepackage{multirow}
\usepackage{siunitx}
\usepackage{amsfonts}
\usepackage{amsmath}
\usepackage{mathrsfs}
\usepackage{mathtools}
\usepackage{amssymb}
\usepackage{amsmath}
\usepackage{graphicx}
\usepackage{graphicx}
\usepackage{epstopdf}
\usepackage{pstricks}
\usepackage{slashed}
\usepackage{mathbbol}
\usepackage{color}
\usepackage{refstyle}
\usepackage{natbib}
\begin{document}
\preprint{OSU-HEP-16-11}
\addtolength{\belowdisplayskip}{-.5ex}
\addtolength{\abovedisplayskip}{-.5ex}
\vspace*{1.2in}

\title{Probing Doubly Charged Higgs Bosons at the LHC through\\[-0.05in] Photon Initiated Processes}

\author{\textbf{K.S. Babu}\footnote{babu@okstate.edu} and \textbf{Sudip Jana}\footnote{sudip.jana@okstate.edu} }
%\email{babu@okstate.edu}
\affiliation{Department of Physics,
Oklahoma State University, Stillwater, OK 74078, USA}

%\author{\textbf{Sudip Jana}}
%\email{sudip.jana@okstate.edu}
%\affiliation{Department of Physics,
%Oklahoma State University, Stillwater, OK 74078-3072, USA}

%\date{\today}

%%%%%%%%%%%%%%%%%%%%%%%%%%%%%%%%%%%%%%%%%%%%%%%%%%%%%%%%%%%%%%%%%%%%%%%%%%%%%%%%%%%%%
\begin{abstract}
\section*{Abstract}
 We show that the photon-photon fusion process contributes significantly to the pair production of doubly charged Higgs bosons at the LHC at a level comparable to the Drell-Yan production. We reinterpret the ATLAS lower limit of 570 GeV (420  GeV) on the mass of $\Delta_{L}^{\pm\pm}$ ($\Delta_{R}^{\pm\pm}$) arising from $SU(2)_L$ triplet (singlet) scalar by including the photon initiated process and derive a new lower limit of 748 GeV (570 GeV), assuming that $\Delta^{\pm \pm}$ decays into $e^\pm e^\pm$ 100\% of the time.   We have also shown that the 5$\sigma$ discovery reach for ${\Delta_{L}^{\pm\pm}}$ (${\Delta_{R}^{\pm\pm}}$) is 846 GeV (783 GeV) with 100 fb$^{-1}$ luminosity  at 13 TeV LHC. We  derive a somewhat more stringent limit on the mass when the doubly charged scalar arises from higher dimensional representations of $SU(2)_L$.
\end{abstract}
%%%%%%%%%%%%%%%%%%%%%%%%%%%%%%%%%%%%%%%%%%%%%%%%%%%%%%%%%%%%%%%%%%%%%%%%%%%%%%%%%%%%%%%%%%%%%%%%%%%%%%%%%%%%%%%%%%%%%%%%%%%%%%%%%%%%%%%%%%%%%%%%%%%%%%%
\maketitle
%%%%%%%%%%%%%%%%%%%%%%%%%%%%%%%%%%%%%%%%%%%%%%%%%%%%%%%%%%%%%%%%%%%%%%%%%%%%%%%%%%%%%%%%%%%%%%%%%%%%%%%%%%%%%%%%%%%%%%%%%%%%%%%%%%%%%%%%%%%%%%%%%%%%%%%

\section{Introduction}

Recently the ATLAS \cite{ATLAS:2014kca,ATLAS:2012hi} and CMS \cite{Chatrchyan:2012ya} collaborations have published results on their searches for doubly charged scalar boson decaying into same sign dileptons.  From the non-observation of any excess compared to the standard model (SM) background, 95$\%$ confidence level (CL) upper limit  on the cross-section and a corresponding lower limit on the mass of the doubly charged scalar boson has been obtained.  The ATLAS collaboration finds a lower limit of 551 GeV on the mass of $\Delta_L^{\pm \pm}$ arising from $SU(2)_L$ triplet, assuming 100\% branching ratio into $e^\pm e^\pm$, with 20.3 $fb^{-1}$ data collected at $\sqrt{s} = 8$ TeV \cite{ATLAS:2014kca}.  The CMS collaboration has quoted an upper limit on the pair production cross section which corresponds to a limit of 382 GeV on the mass of such a $\Delta_L^{\pm \pm}$ obtained with $4.9 fb^{-1}$ data collected at $\sqrt{s}$ = 7 TeV \cite{Chatrchyan:2012ya}.  The ATLAS collaboration has also released its preliminary results obtained with 13.9 $fb^{-1}$ data at $\sqrt{s} = 13$ TeV and quotes an improved lower limit of 570 GeV on the mass of $\Delta_L^{\pm \pm}$ decaying into $e^\pm e^\pm$ \cite{ATLAS:2016pbt}.  These limits have been derived by assuming the pair production of $\Delta_L^{\pm\pm} \Delta_L^{\mp \mp}$ occurs via the Drell-Yan (DY) process (shown in Fig. \ref{1}).  The purpose of this paper is to show the significance of the photo-production process shown in Fig. \ref{2}, which we find to be comparable to the DY process at LHC energies.  We show that by including these photon-initiated processes, the limits on the doubly charged scalar boson can be improved by about 175 GeV, compared to the results quoted by the ATLAS experiment \cite{ATLAS:2016pbt}.

The pair production cross section of $\Delta^{\pm\pm}$ at the LHC strongly depends on the parton luminosities of the proton described by the respective parton distribution functions (PDF). Because of the need for precision phenomenology at the LHC, the PDF of the proton is currently determined using next-to-next-to leading order (NNLO) QCD. At this level of precision, the QED contribution also becomes important. This in particular requires the inclusion of the photon as a parton inside the proton, with an associated distribution function. The NNPDF \cite{Ball:2014uwa,Ball:2013hta}, MRST \cite{Martin:2004dh} and CTEQ \cite{Schmidt:2015zda} collaborations have used different approaches for modeling the photon PDF for the proton.  In our analysis we have adopted the NNPDF approach to describe the photon PDF, which includes the inelastic, semi-elastic and elastic processes (see Fig. \ref{2}), and uses as input the LHC data on Drell-Yan processes. But we have checked that the results are relatively stable when the MRST distribution is used instead.

The photon PDF of the proton at LHC energies was studied in Ref. \cite{Drees:1994zx} for the pair production of charged scalars at the LHC, modeling the PDF theoretically.  This was extended to the study of doubly charged scalars at the LHC arising from $SU(2)_L$ triplet in Ref. \cite{lhctriplet3} which found that the photon fusion process contributed only a fraction $\sim 10\%$ of the DY process.
There is better understanding of the photon PDF of the proton currently, which is less dependent on theoretical modeling.  For a discussion on the theoretical understanding and experimental uncertainties in the PDF extracted from $ep$ scattering data see Ref. \cite{Martin:2014nqa,Manohar:2016nzj}. As a result, we find that the photon fusion process can be as important as the DY process, which enables us to derive improved limits on the doubly charged scalar mass. It should be noted that the photon PDF of the proton has been used in several papers attempting to explain the apparent excess in diphoton invariant mass at 750 GeV (which eventually became statistically insignificant) \cite{diphoton}.  We have checked that our treatment of the photon PDF of the proton indeed reproduces the results of Ref. \cite{diphoton}.

Doubly charge scalar bosons appear in several extensions of the SM.  Type-II sessaw models \cite{seesaw} introduce an $SU(2)_L$ triplet scalar $\Delta_L(1,3,1) = (\Delta_L^{++},\, \Delta_L^+,\,\Delta_L^0)$, where a tiny vacuum expectation value (VEV) of the neutral component $\Delta_L^0$ ($v_{\Delta_L}$) generates small neutrino masses. In left-right symmetric models \cite{lrsm} an $SU(2)_R$ triplet $\Delta_R =(\Delta_R^{++},\, \Delta_R^+,\,\Delta_R^0)$ as well as its parity partner, an $SU(2)_L$ triplet $\Delta_L$, are introduced. The neutral component $\Delta_R^0$ acquires a VEV breaking the $SU(2)_R$ gauge symmetry spontaneously and also generating large Majorana masses for the right-handed neutrinos. The $\Delta_R^{++}$ is a physical field, which is a singlet of $SU(2)_L$, while the $SU(2)_L$ triplet $\Delta_L$ contains a doubly charged scalar $\Delta_L^{\pm \pm}$.
In supersymmetric versions of left-right models \cite{susylrsm}, such doubly charged scalars from $\Delta_R$ survive down to the SUSY breaking scale even when the left-right symmetry is broken at a much higher energy.  Doubly charged scalars also appear in models of radiative neutrino mass models \cite{Zee}, in little Higgs models \cite{modela}, as well as in other extensions of the SM \cite{modelb, modelc}. Collider studies of doubly charged Higgs have been carried out in the context of type-II seesaw  models \cite{lhctriplet1,lhctriplet2, lhctriplet3},  radiative neutrino mass models \cite{lhcrad}, left-right symmetric models \cite{lhclrsm}, little Higgs models \cite{lhclittleh}, and other models \cite{lhcother}. Our main focus in this paper will be $\Delta_L^{\pm \pm}$ arising from an $SU(2)_L$ triplet and $\Delta_R^{\pm \pm}$ which is an $SU(2)_L$ singlet.  These two fields can have direct Yukawa couplings with the leptons ($\Delta_L^{++} \ell_L^- \ell_L^-$ involving left-handed leptons and $\Delta_R^{++} \ell_R^- \ell_R^-$ involving right-handed leptons) and thus are natural candidates for same sign dilepton signatures at the LHC.  We shall also comment briefly on $\Delta^{\pm \pm}$ arising from other $SU(2)_L$ representations.  In this case, however, there must exist additional vector-like leptons to enable couplings with the charged leptons via mixing.

This paper is organized as follows. In Sec. \ref{pd}, we discuss the production and decay modes of $\Delta^{\pm\pm}$. In Sec. \ref{results} we present our analysis methods and the results.  Here we derive improved lower limits on the mass of $\Delta^{\pm \pm}_{L,R}$ and elucidate the discovery reach for a higher luminosity LHC run by including the photon-initiated processes. In Sec. \ref{dch} we analyze the limits when $\Delta^{\pm\pm}$ originates from scalar multiplets other than triplet and singlet of $SU(2)_L$.  Finally we conclude in Sec. \ref{con}.

\section{Production and Decay of Doubly Charged Higgs Boson}\label{pd}

%%%%%%%%%%%%%%%%%%%%%%%%%%%%%%%%%%%%%%%%%%%%%%%%%%%%%%%%%%%%%%%%%%%%%
%%%%%

\begin{figure}[htb!]
$$
 \includegraphics[height=3.5cm, width=7cm]{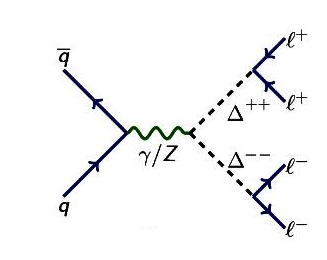}
 $$
 \caption{Feynman diagram for the pair production of $\Delta^{\pm\pm}$ ($p p \rightarrow \Delta^{\pm\pm}\Delta^{\mp\mp}X$)  via Drell-Yan process, with subsequent decays of $\Delta^{\pm \pm}$  into same-sign dileptons.}
\label{1}
\end{figure}
%%%%%%%%%%%%

\begin{figure}[htb!]
$$
 \includegraphics[height=8cm]{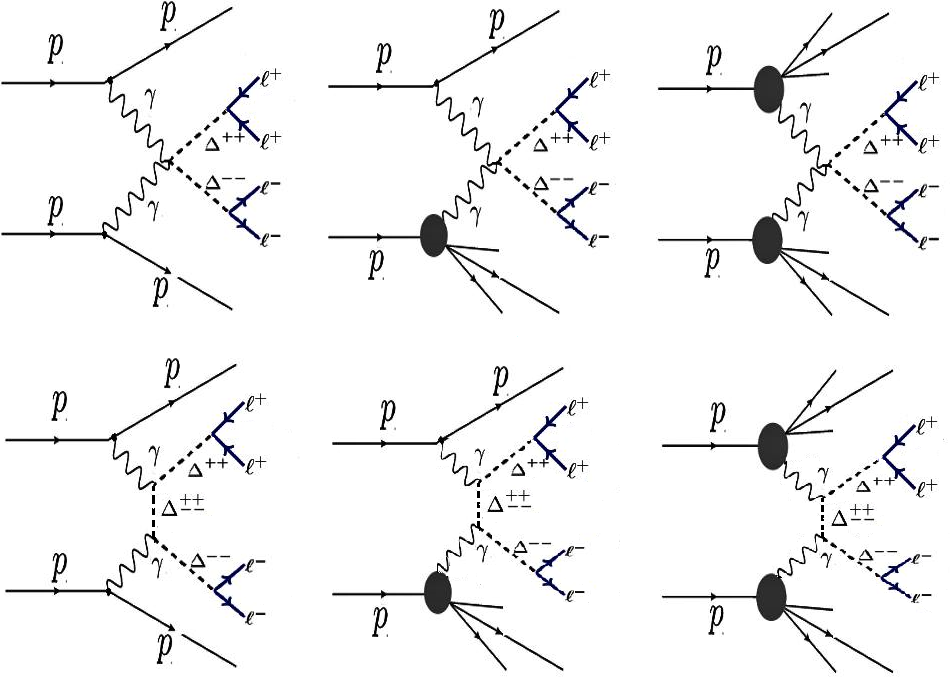}
 $$
 \caption{Feynman diagrams for the pair production of $\Delta^{\pm\pm}$ ($p p \rightarrow \Delta^{\pm\pm}\Delta^{\mp\mp}X$) via photon-photon fusion,
  with subsequent decays of $\Delta^{\pm \pm}$ into same-sign dileptons. Left segment: elastic; middle segment: semi-elastic; and right segment: inelastic scattering sub-processes.}
\label{2}
\end{figure}
%%%%%%%%%%%%%%%%%%%%%

Doubly-charged Higgs bosons can be pair produced  at the LHC via the Drell-Yan (DY) process ($s$-channel photon and $Z$ boson exchange), which is shown in Fig. \ref{1}.  They can also be produced by the photon fusion process shown in Fig. \ref{2}. The ATLAS and CMS collaborations have only kept the DY process in their analyses of doubly charged Higgs boson searches.  As we shall show, the photon fusion process is equally important, and can lead to more stringent limits on the mass of $\Delta^{\pm \pm}$ than the ones quoted by the ATLAS and CMS experiments.

Single production of $\Delta^{++}$ in association with a $W$ boson can occur in the Higgs triplet model; however, this production rate is suppressed by a factor $(v_{\Delta_L}^2/m_W^2)$, where $v_{\Delta_L}$ is the small VEV of $\Delta_L^0$ that generates neutrino masses.  The VEV $v_{\Delta_L}$ is constrained by electroweak $T$ parameter: $v_{\Delta_L} \leq 3$ GeV, and as a result this process is highly suppressed.  Production of $\Delta^{++}$ in association with $\Delta^-$ can occur unsuppressed via the process $u \overline{d} \rightarrow W^{+*} \rightarrow \Delta^{++} \Delta^-$.  However, the signatures of $\Delta^-$ are not very clean, as it decays into final states involving neutrinos. Thus we focus on the pair production of $\Delta^{++} \Delta^{--}$, which would leave clean same sign dilepton signatures in the final state.

The photon fusion channel gets contribution from elastic scattering (where both protons remain intact after the radiation of photons),
semi-elastic scattering (where one of the photons is radiated from the proton, while the other is radiated from the quark parton producing spectator quarks on one side) as well as inelastic scattering (where the two photons are radiated from quark partons of the protons producing spectator quarks on both sides)  as shown in Fig. \ref{2}. The relative contributions of these three processes to the cross section are found to be 4\%, 33\% and 63\% respectively.  We also include the pair production rate through $W$ boson-fusion and $Z$ boson-fusion, but these channels have negligible contributions compared to the photon fusion and DY production channels. The total cross-section from photon photon fusion process ($p(\gamma)p(\gamma)\rightarrow \Delta^{\pm\pm}\Delta^{\mp\mp}$) can be written as \cite{lhctriplet3, Drees:1994zx}:
%%%%%%%%%%%%%
\begin{gather}
\sigma_{\gamma\gamma} = \sigma_{elastic} + \sigma_{inelastic} + \sigma_{semi-elastic}
\\
\sigma_{elastic} = \int_{\tau}^{1} dz_{1}\int_{\tau/z_{1}}^{1} dz_{2}  f_{\gamma/p}(z_{1})f_{\gamma/p^{\prime}}(z_{2})\hat{\sigma}_{\gamma\gamma}(\hat{s} = z_{1} z_{2} s)
\\
\sigma_{semi-elastic} = \int_{\tau}^{1} dx_{1}\int_{\tau/x_{1}}^{1} dz_{1} \int_{\tau/(x_{1}z_{1})}^{1}dz_{2}
\dfrac{1}{x_{1}}F_{2}^{p}(x_{1},Q^{2}) \nonumber \\ f_{\gamma/q}(z_{1})f_{\gamma/p^{\prime}}(z_{2})\hat{\sigma}_{\gamma\gamma}(x_{1} z_{1} z_{2} s)
\\
\sigma_{inelastic} = \int_{\tau}^{1} dx_{1}\int_{\tau/x_{1}}^{1}dx_{2}\int_{\tau/(x_{1}x_{2})}^{1}dz_{1} \int_{\tau/(x_{1}x_{2}z_{1})}^{1}dz_{2} \nonumber \\
\dfrac{1}{x_{1}}F_{2}^{p}(x_{1},Q^{2})\dfrac{1}{x_{2}}F_{2}^{p}(x_{2},Q^{2})f_{\gamma/q}(z_{1})f_{\gamma/q^{\prime}}(z_{2})\hat{\sigma}_{\gamma\gamma}(x_{1} x_{2} z_{1} z_{2} s)
\end{gather}
where $f_{\gamma/p}$ is the photon density inside the proton, $f_{\gamma/q}$ is the photon spectrum inside a quark and $F_{2}^{p}$ is the deep-inelastic proton structure function.
%Approximate expressions for these structure functions are given earlier \cite{lhctriplet3, Drees:1994zx}.
%%%%%%%%%%%%%%%%%%%%%%%

For a doubly charged scalar arising from an arbitrary $SU(2)_L$ multiplet with hypercharge $Y$, the trilinear and quartic gauge interactions relevant for the calculation of the pair production can be written as:
\begin{eqnarray}
\begin{split}
\mathcal{L}^{kin} = \bigg\lbrace i\left[2eA_{\mu}+\dfrac{g}{c_{W}}\left( 2-Y-2s^{2}_{W}\right)Z_{\mu} \right]\Delta^{++}\left(\partial^{\mu}\Delta^{--} \right)
\\
+ \left[2eA_{\mu}+\dfrac{g}{c_{W}}\left( 2-Y-2s^{2}_{W}\right)Z_{\mu}  \right]\times
\left[2eA^{\mu}+\dfrac{g}{c_{W}}\left( 2-Y-2s^{2}_{W}\right)Z^{\mu}  \right]\Delta^{++}\Delta^{--} \bigg\rbrace,
\end{split}
\label{lag}
\end{eqnarray}
where $c_{w}= \cos{\theta_{W}}$, $s_{w}= \sin{\theta_{W}}$, $\theta_{W}$ being the weak mixing angle, and we have used $Q = T_3 + Y$ and set $Q=+2$ for the electric charge of $\Delta^{++}$.  In type-II seesaw models \cite{seesaw}, with an $SU(2)_L$ triplet $\Delta_L$, $Y=+1$, while in the left-right symmetric model \cite{lrsm} there is also an accompanying $\Delta_R^{++}$ with $Y=+2$.  The doubly charged Higgs boson in radiative neutrino mass models \cite{Zee} is analogous to $\Delta_R^{++}$ with $Y=+2$.

%%%%%%%%%%%%%%%%%%%%%%%%%%%%%%%%%%%%%%%%%%%%%%%%%%%%%%%%%%%%%%%%%%%%%%%%
\begin{figure}[htb!]
$$
 \includegraphics[height=7cm, width=0.6\textwidth]{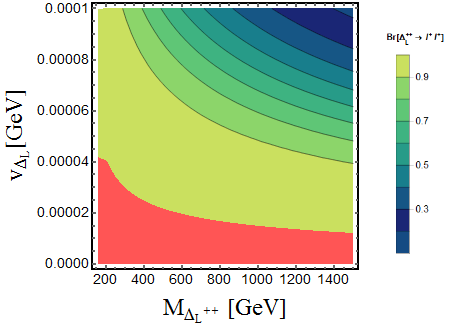}
 $$
 \caption{Contour plot for the branching ratio Br$\left(\Delta_L^{\pm\pm} \rightarrow l^{\pm}l^{\pm}\right)$ in $v_{\Delta_L}$-$M_{\Delta_L^{\pm\pm}}$ plane in the type-II seesaw model. Branching ratio scale is shown on the  right side of the figure. Red shaded zone corresponds to Br$\left(\Delta_L^{\pm\pm} \rightarrow l^{\pm}l^{\pm}\right)=100 \%$.  }
\label{3}
\end{figure}
%%%%%%%%%%%%%%%%%%%%%%%%%%

The doubly charged scalar $\Delta_L^{\pm \pm}$ arising from an $SU(2)_L$ triplet has two primary decays: $\Delta_L^{\pm \pm} \rightarrow \ell_i^\pm \ell_j^\pm$ and $\Delta_L^{\pm\pm} \rightarrow W^\pm W^\pm$.  The widths for these two body decays are given by \cite{lhctriplet2}:
%\item The mass spliting between the triplet members is given by :
%\begin{equation}
%\Delta M^{2} = M^{2}_{\Delta^{\pm}}-M^{2}_{\Delta^{\pm\pm}} = M^{2}_{\Delta^{0}}-M^{2}_{\Delta^{\pm}} = \beta v^{2}/4.
%\end{equation}
\begin{equation}
\Gamma \left(\Delta_L^{\pm\pm} \rightarrow l^{\pm}_{i}l^{\pm}_{j} \right) = \dfrac{\mid M_{\nu}^{ij} \mid ^{2}}{8 \pi (1+\delta_{ij})v_{\Delta_L}^{2}}M_{\Delta_L^{\pm\pm}},
\end{equation}
\begin{equation}
\begin{split}
\Gamma \left(\Delta_L^{\pm\pm} \rightarrow W^{\pm}W^{\pm} \right) = \dfrac{g^{4}v^{2}_{\Delta_L}}{8 \pi M_{\Delta_L^{\pm\pm}}}\sqrt{1-\dfrac{4M^{2}_{W}}{M^{2}_{\Delta_L^{\pm\pm}}}}
\left[2+\left(\dfrac{M^{2}_{\Delta_L^{\pm\pm}}}{2M^{2}_{W}} -1 \right)^{2}\right],
\end{split}
\end{equation}
where $ M_{\nu}^{ij}$ is the $(ij)$ element of the neutrino mass matrix, $\delta_{ij}$ is the Kronecker delta function and $l^{\pm}_{i}= e^{\pm},\mu^{\pm}, \tau^{\pm}$. From these rates it is clear that the branching ratio for $\Delta_L^{\pm\pm}$ decaying into same sign dileptons depends crucially on the triplet VEV $v_{\Delta_L}$.  Taking $ M_{\nu}^{ij}$ to be of order 0.2 eV, for the mass range $M_{\Delta_L^{\pm\pm}} = (200 - 1000)$ GeV, the requirement for the dilepton branching ratio to be dominant is $v_{\Delta_L} \leq 10^{-4}$ GeV.  We shall adopt this constraint, as the decay $\Delta_L^{\pm \pm} \rightarrow W^\pm W^\pm$ is much harder to analyze experimentally owing to large SM background.  A dedicated search for doubly charged Higgs bosons decaying into same sign $W$ boson has not been performed by the ATLAS and CMS collaborations.

It should be noted that $\Delta_L^{\pm \pm}$ can also have a cascade decay as $\Delta_L^{\pm \pm} \rightarrow \Delta_L^\pm W^{\pm *} \rightarrow \Delta_L^0 W^{\pm *} W^{\pm *}$ with the $\Delta_L^0$ decaying into neutrinos, provided that the mass of $\Delta_L^{\pm \pm}$ is larger than that of $\Delta_L^{\pm}$.  The  mass splitting between $\Delta_L^\pm$ and $\Delta_L^{\pm\pm}$ is given by $M_{\Delta_L^\pm}^2 - M_{\Delta_L^{\pm\pm}}^2 = (\beta/4) v^2$, where $\beta$ is a quartic coupling in the Higgs potential \cite{lhctriplet2} and $v = 174$ GeV is the electroweak VEV.  For perturbative values of the coupling $\beta$, the splitting $M_{\Delta_L^\pm}- M_{\Delta_L^{\pm\pm}}$ is only a few tens of GeV.  We shall assume that $\beta > 0$, so that the cascade decay does not proceed.  If $\beta$ were negative, even if the decay $\Delta_L^{\pm \pm} \rightarrow \Delta_L^\pm W^\pm$ is not kinematically allowed for real $W^\pm$, the decay $\Delta_L^{\pm \pm} \rightarrow \Delta_L^\pm \pi^\pm$ will be allowed where a virtual $W^\pm$ boson creates the pion.  Such processes, with nearly degenerate $\Delta_L^{\pm\pm}$ and $\Delta_L^\pm$, will be much more challenging to probe experimentally.

We have shown in Fig. \ref{3} the parameter space of the type-II seesaw model where the same sign dilepton decays of $\Delta^{\pm \pm}$ becomes dominant in the $v_{\Delta_L}-M_{\Delta_L^{\pm\pm}}$ plane.  Here the red shaded region corresponds to nearly 100\% branching ratio into dileptons, and will be the region of interest in our analysis.  This region corresponds to the choice of $v_{\Delta_L} \leq 10^{-4}$ GeV and $\beta >0$ to avoid the cascade decays.

In the case of $\Delta_R^{\pm\pm}$, which is an $SU(2)_L$ singlet field, the decay $\Delta_R^{\pm\pm} \rightarrow W^\pm W^\pm$ does not occur.  The $\Delta_R^{\pm\pm}$ may be accompanied by a $\Delta_R^\pm$ field, as in the radiative neutrino mass model \cite{Zee}.  (Left-right symmetric models also have such $\Delta_R^\pm$ fields, however, this field is part of the Goldstone multiplet associated with the $SU(2)_R$ symmetry breaking.) In this case, the decay $\Delta_R^{\pm \pm} \rightarrow \Delta_R^\pm \Delta_R^\pm$ may occur.  The signature of such decays would have large SM background,  as a result of the neutrino final states arising from the decay of $\Delta_R^\pm$.  In our analysis we assume that the decay $\Delta_R^{\pm \pm} \rightarrow \Delta_R^\pm \Delta_R^\pm$ is not kinematically allowed, so that the dominant decay of $\Delta_R^{\pm\pm}$ is into same sign dileptons.

%%%%%%%%%%%%%%%%%%%%%%%%%%%%%%%%%%%%%%%%%%%%%%%%
\section{Analysis and Results}\label{results}

The ATLAS and CMS collaborations have performed dedicated searches for a doubly charged Higgs boson decaying into same sign dileptons in $pp$ collisions  \cite{ATLAS:2016pbt,ATLAS:2014kca,ATLAS:2012hi,CMS:2016cpz,Chatrchyan:2012ya}.  Lower limits on the mass of the doubly charged scalar have been derived, assuming that the pair production cross section is dominated by the Drell-Yan process. Here we present our results showing the significance of the photon initiated processes, which were ignored in the experimental analyses, and derive improved limits on the mass of $\Delta^{\pm\pm}$.  We also discuss the uncertainties involved in the photon PDF, and project the discovery reach of the LHC for these particles.

%%%%%%%%%%%%%%%%%%%%%%

%%%%%%%%%%%%%%%%%%%%%%%%%%%%%%%%%%%%%%%%%%%%%%%%%%%%%%%%%%%%%%%%%%%%%
\begin{figure}[htb!]
$$
 \includegraphics[height=6cm, width=0.5\textwidth]{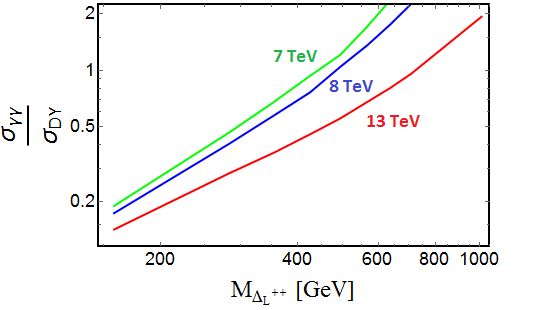}
 \includegraphics[height=6cm, width=0.5\textwidth]{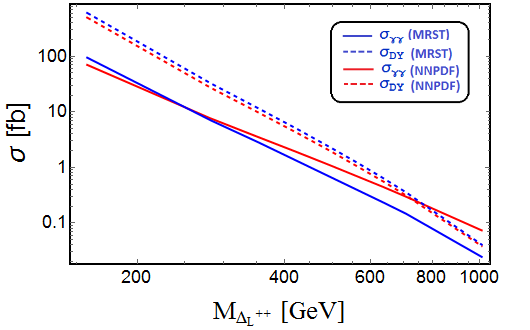}
 $$
 \caption{\textbf{Left}: The ratio of $\sigma_{\gamma\gamma}$ and leading order $\sigma_{DY}$ for doubly charged Higgs pair production at the LHC for different energies using NNPDF parton distribution functions and assuming $\Delta^{\pm\pm}$ belongs to an $SU(2)_L$ triplet. \textbf{Right}: Comparison of DY and photon fusion production cross section for $\Delta_{L}^{\pm\pm}$ at 13 TeV LHC using NNPDF parton distribution functions (red lines) and MRST PDF (blue lines). Dashed line is for DY pair production cross section, whereas solid lines are for pair production cross section via photon fusion.}
\label{4}
\end{figure}
%%%%%%%%%%%%%%%%%%%%%%%%

For our calculations we implement the minimal left right symmetric model (MLRSM)\footnote{This is for our convenience, but we could as well implement other models such as the type-II seesaw  model. Although there will be other channels giving four lepton signals mediated by $Z^\prime$ boson and other heavy neutral Higgs in MLRSM, they are highly suppressed compared to the channel shown in Fig. \ref{1} due to heavy masses of these mediators. The uncertainty due to the presence of these channels in the pair production of $\Delta_L^{\pm\pm}$ is no more than 1$\%$.} in CalcHEP package \cite{Belyaev:2012qa} and we use parton distribution function (PDF) NNPDF23$\char`_$lo$\char`_$as$\char`_$0130$\char`_$qed \cite{Ball:2014uwa,Ball:2013hta}, where the photon PDF\footnote{We can also use MRST2004qed$\char`_$proton \cite{Martin:2004dh}or CT14$\char`_$qedinc \cite{Schmidt:2015zda} where the photon PDF in the proton is inclusive, with the inelastic and elastic contributions included.  Results with MRST2004qed$\char`_$proton \cite{Martin:2004dh} for the pair production cross section is shown in Fig. \ref{4}.} of the proton is inclusive. We calculate the pair production cross-section of $\Delta^{\pm\pm}$ including both the DY and photon fusion processes. The lower limit on the doubly charged Higgs boson pair production cross-section is derived from the experimental analyses \cite{ATLAS:2016pbt,ATLAS:2014kca,Chatrchyan:2012ya} using $\sigma \times$  BR$= N_{rec}/(2\times A \times \epsilon \times \int L dt)$, where $\sigma$ is the pair production cross-section of the doubly charged Higgs $\Delta^{\pm\pm}$, BR is the branching ratio of $\Delta^{\pm\pm}$ decaying into same-sign dileptons, $N_{rec}$ is the number of reconstructed doubly charged Higgs boson candidates, $A\times \epsilon$ is the acceptance times efficiency of the cuts for the respective analyses \cite{ATLAS:2016pbt,ATLAS:2014kca,ATLAS:2012hi,CMS:2016cpz,Chatrchyan:2012ya} and the factor 2 accounts for the two same-sign lepton pairs from the two doubly charged Higgs bosons $\Delta^{++}$ and $\Delta^{--}$. We use the following acceptance criteria: (a) $p_{T}(l)>$ 15 GeV, (b) $\mid{\eta(l)}\mid<$ 2.5 and (c) a veto on any opposite sign dilepton pair invariant mass being close to  the $Z$ boson mass: $\mid M(l^{+}l^{-}) - M_{Z} \mid >$ 15 GeV. The cross-sections and cut efficiencies are estimated by using the CalcHEP package \cite{Belyaev:2012qa}.

We first consider only the DY pair production process, and reproduce the plots shown in the experimental analyses \cite{ATLAS:2016pbt,ATLAS:2014kca,Chatrchyan:2012ya} of the ATLAS and CMS collaborations reasonably well. The QCD correction to this process has been also computed, yielding a next to leading order (NLO) $K$-factor of about 1.25 at the LHC energy for the mass range between 200 GeV and 1 TeV \cite{ATLAS:2016pbt}. The ratio of the two photon contribution relative to the Drell-Yan channel for different LHC energies is shown in Fig. \ref{4}, which clearly shows that the photon fusion process is significant, especially for the higher mass region of $\Delta^{\pm\pm}$.  In the right panel of Fig. \ref{4} we have plotted the pair production cross section with only the DY process included, as well as with only the the photon fusion process included.  Here we show the results for two choices of the PDF, the NNPDF (red lines) and the MRST (blue lines).  We see that the differences in photon fusion cross sections are not much, although it is a bit higher with the use of NNPDF.

%%%%%%%%%%%%%%%%%%%%%%%%%%%%%%%%%%%%%%%%%%%%%%%%%%%%%%%%%%%%%%%%%%%%%%%%
\begin{figure}[htb!]
$$
 \includegraphics[height=6cm, width=0.5\textwidth]{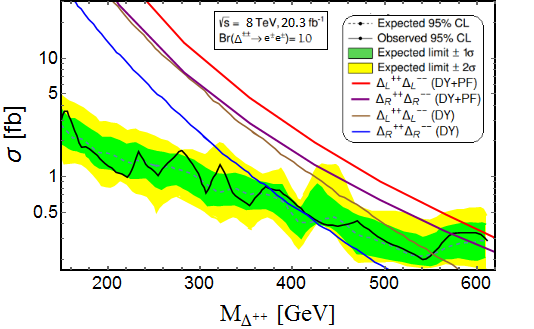}
 \includegraphics[height=6cm, width=0.5\textwidth]{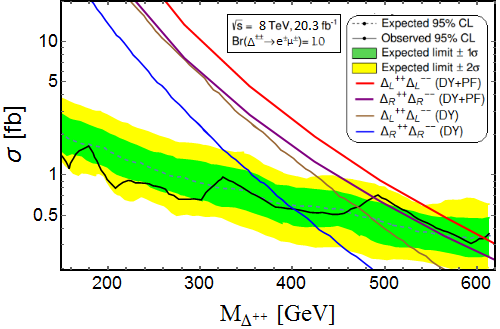}
 $$
 $$
 \includegraphics[height=6cm, width=0.5\textwidth]{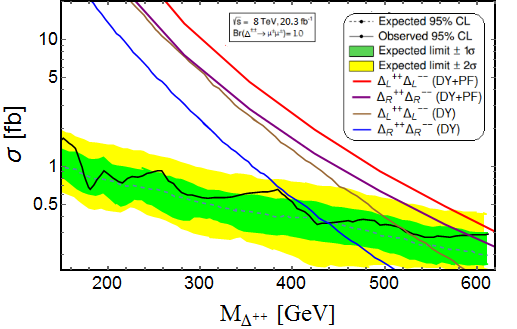}
 $$
 \caption{Upper limits at 95$\%$ C.L. on the cross-section as a function of the dilepton invariant mass for the production of $\Delta^{\pm\pm}$ decaying into (a) $e^{\pm}e^{\pm}$ (top left), (b) $e^{\pm}\mu^{\pm}$ (top right) and (c) $\mu^{\pm}\mu^{\pm}$ pairs (bottom) with a branching ratio  100$\%$ using ATLAS results at $\sqrt{s}$ = 8 TeV with 20.3 fb$^{-1}$ integrated luminosity. The green and yellow regions correspond to the 1$\sigma$ and 2$\sigma$ bands on the expected limits. Red (Brown) solid line is for pair production $pp \rightarrow \Delta^{\pm\pm}_{L}\Delta^{\mp\mp}_{L}$ via Drell-Yan and photon fusion processes (only DY process). Purple (Blue) solid line is for pair production $pp \rightarrow \Delta^{\pm\pm}_{R}\Delta^{\mp\mp}_{R}$ via Drell-Yan and photon fusion processes (only DY process).}
\label{5}
\end{figure}
%%%%%%%%%%%%%%%%%%%%%%%%%%%%%%%%%%%%%%%%%%%%%%%%%%%%%%%%%

%%%%%%%%%%%%

%%%%%%%%%%%%%%%%%%%%%%%%
%%%%%%%%%%%%%%%
The ATLAS collaboration has performed a search \cite{ATLAS:2014kca} for anomalous production of same-sign lepton pairs ($e^{\pm}e^{\pm}, e^{\pm}\mu^{\pm}$ and $\mu^{\pm}\mu^{\pm}$) via pair-produced doubly charged Higgs bosons at the LHC using 20.3 fb$^{-1}$ of data at $\sqrt{s}$ = 8 TeV.  In Fig. \ref{5}, we compare our results with the $\sqrt{s} = 8$ TeV ATLAS results \cite{ATLAS:2014kca}. Upper limits at 95$\%$ C.L. on the cross-section as a function of the like-sign dilepton invariant mass for the production of same-sign lepton pairs ($e^{\pm}e^{\pm}, e^{\pm}\mu^{\pm}$ and $\mu^{\pm}\mu^{\pm}$)  with a branching ratio  100$\%$ are shown in Fig. \ref{5}. The green and yellow regions correspond to the 1$\sigma$ and 2$\sigma$ bands on the expected limits respectively. For this analysis, the ATLAS collaboration did not consider photoproduction.
As a result, the cross section used is significantly smaller than the actual cross section. First we calculated the $\Delta^{\pm\pm}$ pair production cross-section via the DY process. The brown and blue solid lines if Fig. \ref{5} represent the DY pair production cross-section of $\Delta^{\pm\pm}_{L}$ and $\Delta^{\pm\pm}_{R}$ respectively at $\sqrt{s}$ = 8 TeV. According to our DY pair production results, we obtain lower mass limits, assuming a 100$\%$ branching ratio to same-sign dielectrons, of 372 GeV for $\Delta_{R}^{\pm\pm}$ and 551 GeV for $\Delta_{L}^{\pm\pm}$. These limits are almost identical to the ones quoted by the ATLAS collaborations \cite{ATLAS:2014kca}. For other final leptonic states our results agree reasonably well with the ATLAS collaboration results. The solid red (purple) lines in Fig. \ref{5} indicates the pair production cross-section of $\Delta^{\pm\pm}_{L}$ ($\Delta^{\pm\pm}_{R}$)  at $\sqrt{s}$ = 8 TeV considering both DY and photon fusion production mechanisms. After adding the contribution from photon fusion process, 95 $\%$ CL lower mass limits of $\Delta^{\pm\pm}_{L}$ and $\Delta^{\pm\pm}_{R}$ are obtained as 630 GeV and 572 GeV  for 100$\%$ BR to same-sign dielectrons, providing more stringent bounds compared to the ATLAS results based on $\sqrt{s}$ = 8 TeV data.
A summary of the 95$\%$ CL exclusion limits on $M_{\Delta^{\pm\pm}_{L,R}}$ using ATLAS published results at $\sqrt{s}$ = 8 TeV with 20.3 fb$^{-1}$ integrated luminosity is shown in Table \ref{table:2}. Although there are some uncertainties associated with the photon PDF \cite{Ball:2014uwa, Ball:2013hta,Martin:2004dh,Schmidt:2015zda}, our results change only by about 15 GeV or so by using, for example, the MRST photon PDF.

%%%%%%%%%%%%%%%%%%%%%%%%%%%%%%%%%%%%%%%%%%%%%%%%%%%%%%%%%
%%%%%%%%%%%%%%%
\begin{table}[htb!]
\small
  \begin{tabular}{c c c c c c}
    \toprule
    \multirow{2}{*}{\textbf{Benchmark Point}} &
      \multirow{2}{*}{\textbf{ATLAS limit(GeV)}} &
      \multicolumn{3}{c}{\textbf{Limits from our analysis (GeV)}} &
      \\
      &   & \quad \quad \textbf{(DY)} & \quad \quad \textbf{(DY+PF)}\\
      \midrule
    \textbf{$\mathbf{\Delta_{L}^{\pm\pm} \rightarrow e^{\pm}e^{\pm}= 100\%}$} & 551  & \quad \quad 551 & \quad \quad \textbf{$\sim$630} \\
    \textbf{$\mathbf{\Delta_{L}^{\pm\pm} \rightarrow e^{\pm}\mu^{\pm}= 100\%}$} & 468  & \quad \quad 470 & \quad \quad \textbf{\ \ 607}\\
    \textbf{$\mathbf{\Delta_{L}^{\pm\pm} \rightarrow \mu^{\pm}\mu^{\pm}= 100\%}$} & 516  & \quad \quad 515 & \quad \quad \textbf{$\sim$620} \\
     \textbf{$\mathbf{\Delta_{R}^{\pm\pm} \rightarrow e^{\pm}e^{\pm}= 100\%}$} & 374  & \quad \quad 372 & \quad \quad \textbf{\ \ 572} \\
      \textbf{$\mathbf{\Delta_{R}^{\pm\pm} \rightarrow e^{\pm}\mu^{\pm}= 100\%}$} & 402  & \quad \quad 402 & \quad \quad \textbf{\ \ 488} \\
     \textbf{$\mathbf{\Delta_{R}^{\pm\pm} \rightarrow \mu^{\pm}\mu^{\pm}= 100\%}$} & 438 & \quad \quad 439 & \quad \quad \textbf{\ \ 591} \\
    \bottomrule
  \end{tabular}
  \caption{Summary of the 95$\%$ CL exclusion limits on $M_{\Delta^{\pm\pm}_{L,R}}$ using ATLAS results at $\sqrt{s}$ = 8 TeV with 20.3 fb$^{-1}$ integrated luminosity. DY: Drell-Yan pair production; PF: photon fusion process. }
  \label{table:2}
\end{table}
%%%%%%%%%%%%%%%%%%%%%%%%%%%%%%%%%%%%%%%%%%%%%%%%
%%%%%%%%%%%%%%%%%%%%%%%%%%%%%%%%%%%%%%%%%%%%%%%%%%%%%%%%%

%%%%%%%%%%%%%%%%%%%%%%%%%%%%%%%%%%%%%%%%%%%%%%%%%%%%%%%%%%%%%%%%%%%%%%%%
\begin{figure}[htb!]
$$
  \includegraphics[height=6.3cm, width=0.5\textwidth]{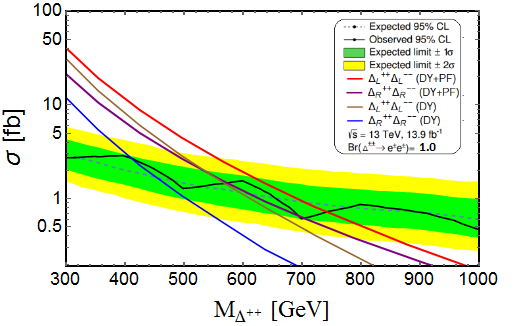}
 \includegraphics[height=6.3cm, width=0.5\textwidth]{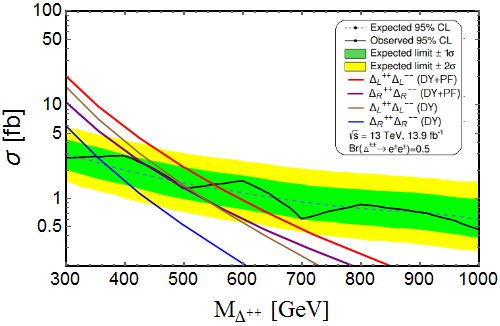}
 $$
 \caption{The observed and expected 95$\%$ C.L. upper limits of the production cross-section \big[$\sigma (p p \rightarrow	\Delta^{\pm\pm}_{L,R} \Delta^{\mp\mp}_{L,R})$\big] as a function of $M_{\Delta^{\pm\pm}_{L,R}}$ using ATLAS results at $\sqrt{s}$ = 13 TeV with 13.9 fb$^{-1}$ integrated luminosity. \textbf{(a) Left:} The limits derived under the assumption that BR($\Delta_{L,R}^{\pm\pm} \rightarrow e^{\pm}e^{\pm}$)= 100$\%$; \textbf{(b) Right:} The limits derived under the assumption that BR($\Delta_{L,R}^{\pm\pm} \rightarrow e^{\pm}e^{\pm}$)= 50$\%$. The green and yellow regions correspond to the 1$\sigma$ and 2$\sigma$ bands on the expected limits respectively. Red (Brown) solid line is for pair production $pp \rightarrow \Delta^{\pm\pm}_{L}\Delta^{\mp\mp}_{L}$ via Drell-Yan and photon fusion processes (only DY process). Purple (Blue) solid line is for pair production $pp \rightarrow \Delta^{\pm\pm}_{R}\Delta^{\mp\mp}_{R}$ via Drell-Yan and photon fusion processes (only DY process).}
\label{6}
\end{figure}
%%%%%%%%%%%%%%%%%%%%%%%%%%%%%%%%%%%%%%%%%%%%%%%%%%
%%%%%%%%%%%%%%%%%%%%%%%%%%%%%%%%%%%%%%%%%%%%%%%%%%%%%%%

A similar search  for doubly charged Higgs bosons decaying into same sign dielectrons has been performed using 13.9 fb$^{-1}$ of $\sqrt{s}$ = 13 TeV pp collision data recorded with the ATLAS detector and preliminary results have been released \cite{ATLAS:2016pbt}. We perform a similar analysis using $\sqrt{s} = 13$ TeV ATLAS results \cite{ATLAS:2014kca} and present our results in Fig. \ref{6}. In Ref. \cite{ATLAS:2016pbt}, it is clearly stated that the production of $\Delta^{\pm\pm}$ was allowed only via the DY process during signal processing. For pair production of $\Delta^{\pm\pm}$, the lower bounds \cite{ATLAS:2016pbt} on the $\Delta_{L}^{\pm\pm}$ ($\Delta_{R}^{\pm\pm}$) mass are set 570 and 530 GeV  (420 and 380 GeV) in the 100$\%$ and 50$\%$ branching fraction scenarios for  final leptonic states $e^{\pm}e^{\pm}$  by the ATLAS collaborations. Our analysis reproduces these results when only the DY process is included.  From a full analysis including pair production via both DY process and photon fusion process,  95 $\%$ CL lower mass limits of $\Delta^{\pm\pm}_{L}$ and $\Delta^{\pm\pm}_{R}$ are obtained as 748 GeV (554 GeV) and 570 GeV (516 GeV) for 100$\%$ (50$\%$) BR to same-sign dielectrons, providing more stringent bounds compared to the preliminary ATLAS results. Our results are  summarized in Table \ref{table:1}.

%%%%%%%%%%%%%%%

\begin{table}[htb!]
\small
  \begin{tabular}{c c c c c c}
    \toprule
    \multirow{2}{*}{\textbf{Benchmark Point}} &
      \multirow{2}{*}{\textbf{ATLAS limit(GeV)}} &
      \multicolumn{3}{c}{\textbf{Limits from our analysis (GeV)}} &
      \\
      &   & \quad \quad \textbf{(DY)} & \quad \quad \textbf{(DY+PF)}\\
      \midrule
    \textbf{$\mathbf{\Delta_{L}^{\pm\pm} \rightarrow e^{\pm}e^{\pm}= 100\%}$} & 570  & \quad \quad 569 & \quad \quad \textbf{748} \\
    \textbf{$\mathbf{\Delta_{L}^{\pm\pm} \rightarrow e^{\pm}e^{\pm}= 50\%}$} & 530  & \quad \quad 524 & \quad \quad \textbf{554}\\
    \textbf{$\mathbf{\Delta_{R}^{\pm\pm} \rightarrow e^{\pm}e^{\pm}= 100\%}$} & 420  & \quad \quad 418 & \quad \quad\textbf{570} \\
   \textbf{$\mathbf{\Delta_{R}^{\pm\pm} \rightarrow e^{\pm}e^{\pm}= 50\%}$} & 380 & \quad \quad 377 & \quad \quad \textbf{516} \\
     \bottomrule
  \end{tabular}
  \caption{Summary of the 95$\%$ CL exclusion limits on $M_{\Delta^{\pm\pm}_{L,R}}$ using ATLAS results at $\sqrt{s}$ = 13 TeV with 13.9 fb$^{-1}$ integrated luminosity. DY: Drell-Yan pair production; PF: photon fusion process. }
  \label{table:1}
\end{table}
%%%%%%%%%%%%%%%%%%%%%%%%%%%%%%%%%%%%%%%%%%%%%%%%%%%%

%%%%%%%%%%%%%%%%%%%%%%%%%%%%%%%%%%%%%%%%%%%%%%%%%%%%%%%%%%%%%%%%%%%%%%%%
\begin{figure}[htb!]
$$
 \includegraphics[height=6.1cm, width=0.5\textwidth]{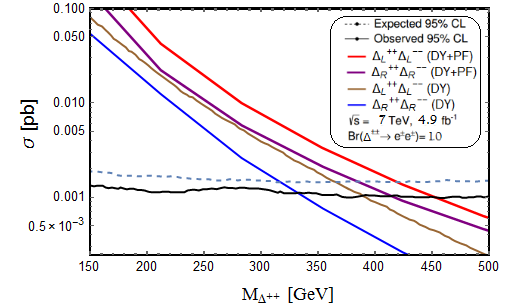}
 \includegraphics[height=6.1cm, width=0.5\textwidth]{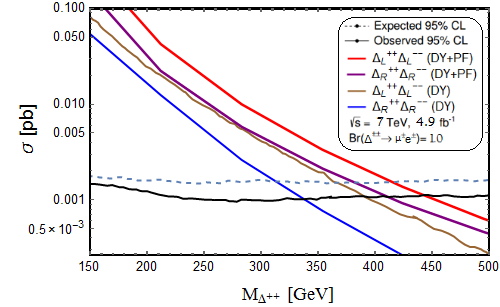}
 $$
 $$
 \includegraphics[height=6.1cm, width=0.5\textwidth]{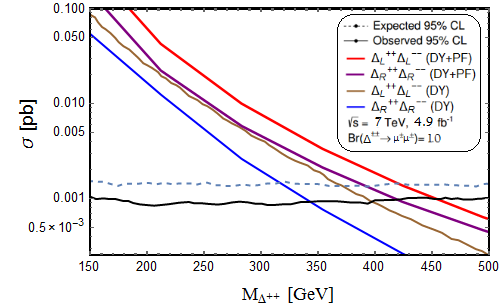}
 $$
 \caption{Upper limits at 95$\%$ C.L. on the cross-section as a function of the dilepton invariant mass for the production of $\Delta^{\pm\pm}$ decaying into (a) $e^{\pm}e^{\pm}$ (top left), (b) $e^{\pm}\mu^{\pm}$ (top right) and (c) $\mu^{\pm}\mu^{\pm}$ pairs (bottom) with a branching ratio  100$\%$ using CMS results at $\sqrt{s}$ = 7 TeV with 4.9 fb$^{-1}$ integrated luminosity. Red (Brown) solid line is for pair production $pp \rightarrow \Delta^{\pm\pm}_{L}\Delta^{\mp\mp}_{L}$ via Drell-Yan and photon fusion processes (only DY process). Purple (Blue) solid line is for pair production $pp \rightarrow \Delta^{\pm\pm}_{R}\Delta^{\mp\mp}_{R}$ via Drell-Yan and photon fusion processes (only DY process).}
\label{7}
\end{figure}
%%%%%%%%%%%%%%%%%%%%%%%%%%%%%%%%%%%%%%%%%%%%%%%%%%%%%%%%%

 We have also done a reanalysis of the CMS results \cite{Chatrchyan:2012ya} at $\sqrt{s}$ = 7 TeV with 4.9 fb$^{-1}$ integrated luminosity. Here  also we find more stringent upper limits on the cross section for $\Delta^{\pm\pm}$ pair production with the inclusion of the photon fusion contribution.
 Although CMS collaboration did not set any bound on $\Delta_{R}^{\pm\pm}$ mass, we derive mass limits in both situations -- DY only included, and DY plus photon fusion processes included. In Fig. \ref{7} we plot the CMS results at $\sqrt{s} = 7$ TeV on the cross section and the invariant mass of $\Delta_{L,R}^{\pm\pm}$ for various scenarios as noted in the figure caption for the branching ratios.  
 Our improved bounds are summarized in Table \ref{table:3}. The most stringent lower mass limit of $\Delta_{L}^{\pm\pm}$ ($\Delta_{R}^{\pm\pm}$) is found to be 453 GeV (397 GeV) at  of $95\%$ CL, with the assumption that Br($\Delta_{L,R}^{\pm\pm} \rightarrow \mu^{\pm}\mu^{\pm}$)= 100$\%$, providing significantly more stringent constraints than previously published limits.
%%%%%%%%%%%%%%%
\begin{table}[htb!]
\small
  \begin{tabular}{c c c c c c}
    \toprule
    \multirow{2}{*}{\textbf{Benchmark Point}} &
      \multirow{2}{*}{\textbf{CMS limit(GeV)}} &
      \multicolumn{3}{c}{\textbf{Limits from our analysis (GeV)}} &
      \\
      &   & \quad \quad \textbf{(DY)} & \quad \quad \textbf{(DY+PF)}\\
      \midrule
    \textbf{$\mathbf{\Delta_{L}^{\pm\pm} \rightarrow e^{\pm}e^{\pm}= 100\%}$} &  382  & \quad \quad 387 & \quad \quad \textbf{452} \\
    \textbf{$\mathbf{\Delta_{L}^{\pm\pm} \rightarrow e^{\pm}\mu^{\pm}= 100\%}$} &  391 &  \quad \quad 392 & \quad \quad \textbf{442}\\
    \textbf{$\mathbf{\Delta_{L}^{\pm\pm} \rightarrow \mu^{\pm}\mu^{\pm}= 100\%}$} &  395  & \quad \quad 397 & \quad \quad \textbf{453} \\
     \textbf{$\mathbf{\Delta_{R}^{\pm\pm} \rightarrow e^{\pm}e^{\pm}=100\%}$} & \textemdash & \quad \quad 329 & \quad \quad \textbf{414} \\
      \textbf{$\mathbf{\Delta_{R}^{\pm\pm} \rightarrow e^{\pm}\mu^{\pm}= 100\%}$} & \textemdash & \quad \quad 336 & \quad \quad \textbf{410} \\
     \textbf{$\mathbf{\Delta_{R}^{\pm\pm} \rightarrow \mu^{\pm}\mu^{\pm}= 100\%}$} & \textemdash & \quad \quad 342 &  \quad \quad  \textbf{420} \\
   \bottomrule
  \end{tabular}
  \caption{Summary of the 95$\%$ CL exclusion limits on $M_{\Delta^{\pm\pm}_{L,R}}$ using CMS results at $\sqrt{s}$ = 7 TeV with 4.9 fb$^{-1}$ integrated luminosity. DY: Drell-Yan pair production; PF: photon fusion process. }
  \label{table:3}
\end{table}
%%%%%%%%%%%%%%%%%%%%%%%%%%%
%%%%%%%%%%%%%%%%%%%%%%%
\begin{figure}[htb!]
$$
\includegraphics[height=6.5cm]{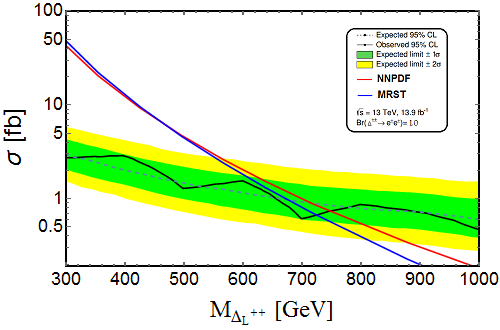}
 $$
 \caption{95$\%$ CL exclusion limits on $M_{\Delta^{\pm\pm}_{L,}}$ using ATLAS results at $\sqrt{s}$ = 13 TeV with 13.9 fb$^{-1}$ integrated luminosity. Black solid line: Observed limit; Blue dotted line: Expected limit; Red solid line: Production cross-section considering both DY and photon fusion processes using parton distribution function NNPDF; Blue solid line: Production cross-section considering both DY and photon fusion processes using parton distribution function MRST. The limit is derived under assumption that BR to same-sign dielectrons is 100$\%$. }
\label{8}
\end{figure}
%%%%%%%%%%%%%%%%%%%%
%%%%%%%%%%%%%%%%%%%%%%%%%%%%%%%%%%%%%%%%%%%%%%%%%%%%%%%%%%%%%%%%%%%%
\begin{figure}[htb!]
$$
 \includegraphics[height=6cm, width=0.5\textwidth]{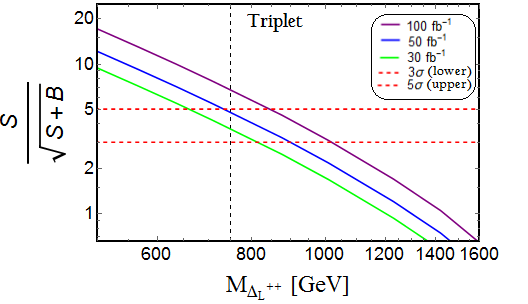}
 \includegraphics[height=6cm, width=0.5\textwidth]{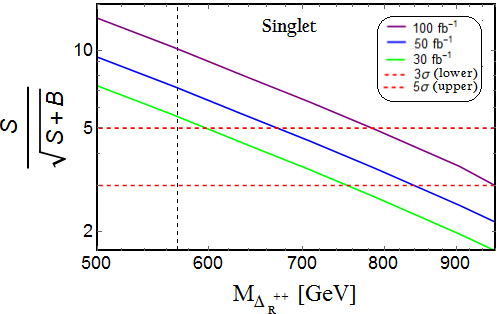}
 $$
 \caption{\textbf{Left:} Significance versus M$_{\Delta_L^{\pm\pm}}$ plot assuming BR($\Delta_L^{\pm\pm} \rightarrow l^{\pm}l^{\pm}$) = 100$\%$ at 13 TeV LHC for 30 fb$^{-1}$, 50 fb$^{-1}$   and 100 fb$^{-1}$ luminosities. Left part of dashed black line is excluded by the current experimental limit as derived earlier. Here doubly charged scalar is  from an $SU(2)_L$ triplet.\\
 \textbf{Right :} Significance versus M$_{\Delta_R^{\pm\pm}}$ plot assuming BR($\Delta_R^{\pm\pm} \rightarrow l^{\pm}l^{\pm}$) = 100$\%$ at 13 TeV LHC for 30 fb$^{-1}$, 50 fb$^{-1}$   and 100 fb$^{-1}$ luminosities. Left part of dashed black line is excluded by the current experimental limit as derived earlier. Here the doubly charged scalar is an $SU(2)_L$ singlet.}
\label{9}
\end{figure}
%%%%%%%%%%%%%%%%%%%%

 As noted in Ref. \cite{Ball:2014uwa}, while using NNPDF 2.3QED PDF set, for invariant mass $M_{ll}$ above $M_{Z}$, corrections due to PDF uncertainties become sizable, more than a few percent and up to 20$\%$ for  very mass high values.   Taking the worst case scenario of 20\% uncertainty in the PDF, we find that the lower mass limit decreases by about 18 GeV, which is still much stronger than the limit of 570 GeV derived in reference \cite{ATLAS:2016pbt} using 13 TeV $pp$ collision data recorded with the ATLAS detector with 13.9 fb$^{-1}$ data.
 In Fig. \ref{8}, we show the variation on the lower mass limit with respect to changing the PDF.  Here we have plotted the mass limits using the  MRST and the NNPDF distribution functions. We obtain a slightly relaxed  lower limit of 729.6 GeV  with the MRST PDF on M$_{\Delta^{\pm\pm}_{L}}$, which differs from the value of 748 GeV obtained with the NNPDF only by 18 GeV. We conclude that the lower limits derived by including the photon fusion process is rather stable and reliable under the change of the PDF.
%%%%%%%%%%%%%%%%%%%%%%%%%%%%%%%%%%%%%%%%%%%%%%%%%%%%%%%%%%

Now we analyze the discovery reach  of $\Delta^{\pm\pm}$ for higher luminosities at the 13 TeV LHC in the four lepton signal
from the decays of $\Delta^{\pm\pm}\rightarrow l^{\pm}l^{\pm}$. After employing the previously mentioned cuts for the events, these  signal events would have almost no SM background.  If we reconstruct the invariant mass for same-sign dileptons, it will give a sharp peak at $M_{\Delta^{\pm\pm}}$ with no SM background. Here we choose BR($\Delta^{\pm\pm} \rightarrow l^{\pm}l^{\pm}$) = 100$\%$. The significance ($S/\sqrt{S+B}$) has been plotted in Fig. \ref{9} as a function of $M_{\Delta^{\pm\pm}}$ for near-future LHC luminosities of 30 fb$^{-1}$, 50 fb$^{-1}$   and 100 fb$^{-1}$. We have found that at 5$\sigma$ level the $M_{\Delta_L^{\pm\pm}}$ ($M_{\Delta_R^{\pm\pm}}$) can be probed up to 846 GeV (783 GeV) for 100 fb$^{-1}$ luminosity, 735 GeV (670 GeV) for 50 fb$^{-1}$ luminosity and 658 GeV (597 GeV) for 30 fb$^{-1}$ luminosity.
%%%%%%%%%%%%%%%%%%%%%%%%%%%%%%%%%%%%%%%%%%%%%%%%%%%%%%%%%%%%%%%%%%%%%%

%%%%%%%%%%%%%%%%%%%%%%%%%%%%%%%%%%%%%%%%%%%%%%%%%%%%%%%%%%%%%%%%%%%%%%%%
%%%%%%

\begin{table}[htb!]
\small
  \begin{tabular}{c c c c c}
    \toprule
    \multirow{2}{*}{\textbf{Benchmark Point}} &
      \multicolumn{2}{c}{\textbf{M$_{\Delta_L^{\pm\pm}}$ [GeV]}} &
      \multicolumn{2}{c}{\textbf{M$_{\Delta_R^{\pm\pm}}$ [GeV]}} \\
      & {(3$\sigma$ limit)} & {(5$\sigma$ limit)} & {(3$\sigma$ limit)} & {(5$\sigma$ limit)}\\
      \midrule
   \textbf{$\textit{l}$ = 30 fb$^{-1}$} & 812 & 658 & 750 & 597 \\
  \textbf{$\textit{l}$ = 50 fb$^{-1}$} & 900 & 735 & 838 &  670 \\
   \textbf{$\textit{l}$ = 100 fb$^{-1}$} & 1020 & 846 & 957 & 783  \\
    \bottomrule
  \end{tabular}
   \caption{Summary of $\Delta_{L,R}^{\pm\pm}$ mass reach at the 13 TeV LHC. Here \textit{l} = luminosity. }
  \label{table:reach}
\end{table}
%%%%%%%%%%%%%%%%%%%%%%%%%%%%%%%%%%%%%%%%%%%%%%%%%%%%%%%%%%%%%%%%%%%%%%%%%%%%%%%%%%%%%%%%%%%%
%%%%%%%%%%%%%%
\section{Doubly Charged Higgs from Different \boldmath${SU(2)_L}$ Multiplets}\label{dch}
%%%%%%%%%%%%%%%%%%%%%%%%%%%%%%%%%%%%%%%%%%%%%%%%%%%%%%%%%%%%%%%%
\begin{figure}[htb!]
 $$
\includegraphics[height=6cm]{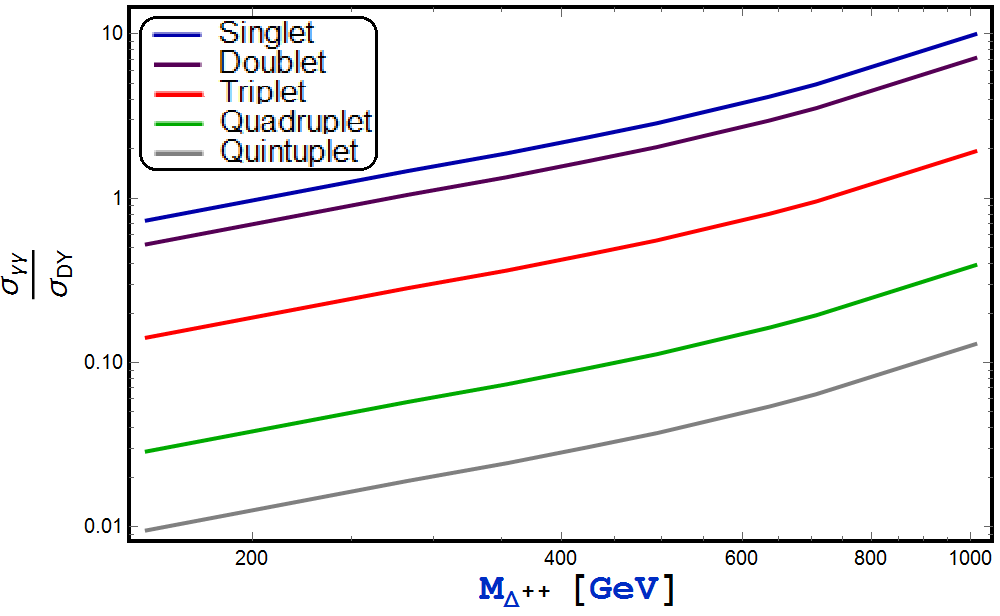}
 $$
 \caption{ The ratio between $\sigma_{\gamma\gamma}$ and leading order $\sigma_{DY}$ for doubly charged Higgs pair production at the 13 TeV LHC for different choice of $SU(2)_L$ multiplets. From top to bottom, $\Delta^{\pm\pm}$ belongs to  singlet (blue), doublet (purple), triplet (red), quadruplet (green) and quintuplet (gray).}
\label{multipletratio}
\label{10}
\end{figure}
%%%%%%%%%%%%%%%%%%%%%%%%
%%%%%%%%%%%%%%%%%%%%%%%%%%%%%%%%%%%%%%%%%%%%%%%%%%%%%%%%%%%%%%%%%%%%
\begin{figure}[htb!]
$$
  \includegraphics[height=6.5cm]{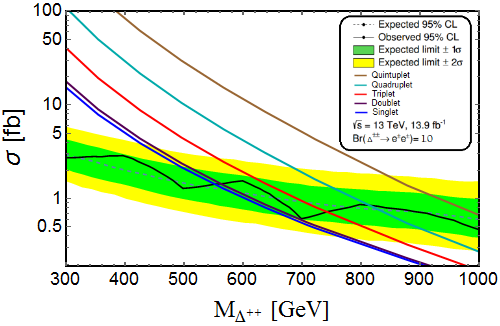}
 $$
 \caption{95$\%$ CL exclusion limits on $M_{\Delta^{\pm\pm}_{L,}}$ using ATLAS results at $\sqrt{s}$ = 13 TeV with 13.9 fb$^{-1}$ integrated luminosity. Black solid line: Observed limit; Blue dotted line: Expected limit. From top to bottom, brown, cyan, red, purple and blue solid lines are model predicted cross sections, when $\Delta^{\pm\pm}$ belongs to quintuplet, quadruplet, triplet, doublet and singlet respectively. The limit is derived under assumption that BR to same-sign dielectrons is 100$\%$. }
\label{y}
\end{figure}
%%%%%%%%%%%%%%%%%%%%
In previous sections, we mainly focused on $\Delta^{\pm\pm}$ arising from a $SU(2)_L$ triplet or a singlet. Such states can have direct couplings to two leptons.  Now we generalize and analyze cases where $\Delta^{\pm\pm}$ originates from a different $SU(2)_L$ multiplet. We assume that its decay is dominantly into same sign dileptons.  This would require the existence of vector-like leptons, which can mix with the ordinary leptons and facilitate such decays. For illustration purposes we restrict ourselves to the cases where the $\Delta^{++}$ has the maximal electric charge in the multiplet.   We allow the following representations under $SU(2)_{L}$:
\begin{itemize}
\item$\Delta^{++}$ in a singlet :
$\phi = \Delta^{++}$; ($T=0, T_{3}=0, Y=2$).
\item$\Delta^{++}$ in a doublet :
 $\phi = (\Delta^{++},\Delta^{+})$; ($T=1/2, T_{3}=1/2, Y=3/2$).
\item$\Delta^{++}$ in a triplet :
$\phi = (\Delta^{++},\Delta^{+}, \Delta^{0})$; ($T=1, T_{3}=1, Y=1$).
\item$\Delta^{++}$ in a quadruplet :
 $\phi = (\Delta^{++},\Delta^{+}, \Delta^{0}, \Delta^{\prime^{-}})$; ($T=3/2,T_{3}=3/2, Y=1/2$).
\item$\Delta^{++}$ in a quintuplet : $\phi$ = $(\Delta^{++},\Delta^{+}, \Delta^{0}, \Delta^{-}, \Delta^{--})$; ($T=2, T_{3}=2, Y=0$).
\end{itemize}
Here the electric charge is defined as Q = $T_{3}+Y$, where $T_{3}$ is the third component of isospin and $Y$ is the hypercharge, and the relevant gauge interactions are given in Eq. (\ref{lag}).
In Fig.  \ref{multipletratio}, we have shown the ratio between $\sigma_{\gamma\gamma}$ and leading order $\sigma_{DY}$ for doubly charged Higgs pair production at the 13 TeV LHC for different choices of the multiplets. From this plot we see that for the higher mass region of $\Delta^{\pm\pm}$, photon photon fusion contribution becomes much more significant compared to the DY process. Due to the different $Z\Delta^{\pm\pm}\Delta^{\mp\mp}$ couplings for different multiplets (-0.33 for singlet, 0.029 for doublet, 0.388 for triplet, 0.747 for quadruplet and 1.106 for quintuplet), DY pair production rate increases successively from singlet to quintuplet, whereas due to the indifferent couplings $\gamma\Delta^{\pm\pm}\Delta^{\mp\mp}$ and $\gamma\gamma\Delta^{\pm\pm}\Delta^{\mp\mp}$,  the pair production rate via photon fusion process will remain the same. As a result, the ratio between $\sigma_{\gamma\gamma}$ and $\sigma_{DY}$ for doubly charged Higgs pair production at the LHC will decrease from singlet to quintuplet successively, as shown in Fig. \ref{multipletratio}. Now we derive the lower mass limits for each cases using $\sqrt{s} = 13 $TeV ATLAS results \cite{ATLAS:2016pbt}, which is shown in Fig. \ref{y}. The mass bounds on $\Delta^{\pm\pm}$ for different multipets are summarized in Table \ref{table:y} using 13.9 fb$^{-1}$ of $\sqrt{s}=$ 13 TeV ATLAS data at $95\%$ CL, with the assumption that Br($\Delta^{\pm\pm} \rightarrow e^{\pm}e^{\pm}$)= 100$\%$.

%%%%%%%%%%%%%%%
\begin{table}[htb!]
  \begin{tabular}{c c c}
    \toprule
    \multirow{2}{*}{\textbf{Multiplet}} &
      \multirow{2}{*}{{$\mathbf{\Delta^{\pm\pm}}$} \textbf{Mass Limit [GeV]}} &
      \\ \\
    \midrule
   \textbf{Singlet} & \  \ \textbf{570} \\
   \textbf{Doublet} & \  \ \textbf{577} \\
   \textbf{Triplet} & \  \ \textbf{748} \\
   \textbf{Quadruplet} & \ \ \textbf{813} \\
   \textbf{Quintuplet} & $\sim$ \textbf{1100}
    \\
    \bottomrule
  \end{tabular}
  \caption{Summary of the 95$\%$ CL exclusion limits on $M_{\Delta^{\pm\pm}}$ using ATLAS results at $\sqrt{s}$ = 13 TeV with 13.9 fb$^{-1}$ integrated luminosity for different choices of  $SU(2)_L$ multiplets. These limits are derived under the assumption that BR to same-sign dielectrons is 100$\%$.}
  \label{table:y}
\end{table}
%%%%%%%%%%%%%%%%%%%%%%%%%%%%%%%%%%%%%%%%%%%%%%%%%%%%%%%%%

\section{Summary and Discussions}\label{con}
In this paper we have reinvestigated the pair production of doubly charged scalars at the LHC. Pair-production, in spite of its relative kinematical suppression, has the advantage of being relatively model independent. We have found that the photon fusion process, which has been neglected in the experimental analyses thus far, contributes to the pair production cross section at a level comparable to the Drell-Yan production process.  We focused on the most spectacular four lepton final state originating from the decays of the $\Delta^{\pm\pm}$ into same sign lepton pairs. These channels not only lead to remarkably background-free signatures of the doubly charged scalars, but they also demonstrate a crucial link between observations at high energy colliders and widely discussed mechanisms of neutrino mass generation.

By including the photon fusion process in the production cross section, we are able to derive more stringent lower mass limits on $\Delta^{\pm\pm}$ than previously quoted.  First we reproduced the limits quoted by the ATLAS and CMS collaborations by only including the DY production sub-process.  With the photon fusion process included, we have derived, from the same data, more stringent limits on the mass of $\Delta^{\pm\pm}$.  These  results are summarized in Table \ref{table:1}, Table \ref{table:2} and Table \ref{table:3}, corresponding to data analyzed by the ATLAS collaboration at $\sqrt{s} = 8$ TeV, $\sqrt{s} = 13$ TeV and by the CMS collaboration at $\sqrt{s} = 7$ TeV. These results represent a significant improvement over previous ATLAS and CMS results. We have analyzed the discovery reach for $\Delta^{\pm\pm}_{L,R}$ in the upcoming run at the LHC, which are  shown in Table \ref{table:reach}. We have also shown in Table \ref{table:y} the different mass limits for $\Delta^{\pm\pm}$ belonging to different types of $SU(2)_L$ multiplets.
%%%%%%%%%%%%%%%%%%%%%%%%%%%%%%%%%%%%%%%%%%%%%%%%%%%%%%%%%%%%%%%%%%%%%%%%

\section*{Acknowledgement}
We thank A. Khanov, K. Ghosh, T. Ghosh and A. Pukhov for useful discussions. This work is supported in part by the US Department of Energy Grant No. de-sc0016013.

%%%%%%%%%%%%%%%%%%%%%%%%%%%%%%%%%%%%%%%%%%%
%%%%%%%%%%%%%%%%%%%%%%%%%%%%%%%%%%%%%%%%%%%%%%%%%%%%%%%%%%%%%%%%%%%%%%%%%%%%%


\begin{thebibliography}{99}
\section*{References}

\bibitem{ATLAS:2014kca}
  G.~Aad {\it et al.} [ATLAS Collaboration],
  %``Search for anomalous production of prompt same-sign lepton pairs and pair-produced doubly charged Higgs bosons with $ \sqrt{s}=8 $ TeV $pp$ collisions using the ATLAS detector,''
  JHEP {\bf 1503}, 041 (2015)
  %doi:10.1007/JHEP03(2015)041
  %[arXiv:1412.0237 [hep-ex]].
  %%CITATION = doi:10.1007/JHEP03(2015)041;%%
  %68 citations counted in INSPIRE as of 29 Oct 2016

%\cite{ATLAS:2012hi}
\bibitem{ATLAS:2012hi}
  G.~Aad {\it et al.} [ATLAS Collaboration],
  %``Search for doubly-charged Higgs bosons in like-sign dilepton final states at $\sqrt{s}=7$ TeV with the ATLAS detector,''
  Eur.Phys.J.C {\bf 72}, 2244 (2012)
  %doi:10.1140/epjc/s10052-012-2244-2
 % [arXiv:1210.5070 [hep-ex]].
  %%CITATION = doi:10.1140/epjc/s10052-012-2244-2;%%
  %120 citations counted in INSPIRE as of 29 Oct 2016


%\cite{Chatrchyan:2012ya}
\bibitem{Chatrchyan:2012ya}
  S.~Chatrchyan {\it et al.} [CMS Collaboration],
  %``A search for a doubly-charged Higgs boson in $pp$ collisions at $\sqrt{s}=7$ TeV,''
  Eur.Phys.J.C {\bf 72}, 2189 (2012)
  %doi:10.1140/epjc/s10052-012-2189-5
  %[arXiv:1207.2666 [hep-ex]].
  %%CITATION = doi:10.1140/epjc/s10052-012-2189-5;%%
  %140 citations counted in INSPIRE as of 29 Oct 2016

%\cite{ATLAS:2016pbt}
\bibitem{ATLAS:2016pbt}
The ATLAS collaboration [ATLAS Collaboration],
  %``Search for doubly-charged Higgs bosons in same-charge electron pair final states using proton-proton collisions at $\sqrt{s}=13\,\mathrm{TeV}$ with the ATLAS detector,''
  ATLAS-CONF-2016-051.
  %%CITATION = ATLAS-CONF-2016-051;%%
  %2 citations counted in INSPIRE as of 29 Oct 2016

 %\cite{Ball:2014uwa}
\bibitem{Ball:2014uwa}
  R.~D.~Ball {\it et al.} [NNPDF Collaboration],
  %``Parton distributions for the LHC Run II,''
  JHEP {\bf 1504}, 040 (2015)
  %doi:10.1007/JHEP04(2015)040
%  [arXiv:1410.8849 [hep-ph]].
  %%CITATION = doi:10.1007/JHEP04(2015)040;%%
  %461 citations counted in INSPIRE as of 31 Oct 2016

  %\cite{Ball:2013hta}
\bibitem{Ball:2013hta}
  R.~D.~Ball {\it et al.} [NNPDF Collaboration],
  %``Parton distributions with QED corrections,''
  Nucl.Phys.B {\bf 877}, 290 (2013).
  %doi:10.1016/j.nuclphysb.2013.10.010
%  [arXiv:1308.0598 [hep-ph]].
  %%CITATION = doi:10.1016/j.nuclphysb.2013.10.010;%%
  %189 citations counted in INSPIRE as of 31 Oct 2016


%\cite{Martin:2004dh}
\bibitem{Martin:2004dh}
  A.~D.~Martin, R.~G.~Roberts, W.~J.~Stirling and R.~S.~Thorne,
  %``Parton distributions incorporating QED contributions,''
  Eur.Phys.J.C {\bf 39}, 155 (2005)
  %doi:10.1140/epjc/s2004-02088-7
%  [hep-ph/0411040].
  %%CITATION = doi:10.1140/epjc/s2004-02088-7;%%
  %354 citations counted in INSPIRE as of 31 Oct 2016

%\cite{Schmidt:2015zda}
\bibitem{Schmidt:2015zda}
  C.~Schmidt, J.~Pumplin, D.~Stump and C.~P.~Yuan,
  %``CT14QED parton distribution functions from isolated photon production in deep inelastic scattering,''
  Phys.Rev.D{\bf 93}, no. 11, 114015 (2016)
  %doi:10.1103/PhysRevD.93.114015
  %[arXiv:1509.02905 [hep-ph]].
  %%CITATION = doi:10.1103/PhysRevD.93.114015;%%

%\cite{Drees:1994zx}
\bibitem{Drees:1994zx}
  M.~Drees, R.~M.~Godbole, M.~Nowakowski and S.~D.~Rindani,
  %``gamma gamma processes at high-energy p p colliders,''
  Phys.\ Rev.\ D {\bf 50}, 2335 (1994)
%  doi:10.1103/PhysRevD.50.2335
%  [hep-ph/9403368].
  %%CITATION = doi:10.1103/PhysRevD.50.2335;%%
  %62 citations counted in INSPIRE as of 28 Dec 2016

  %\cite{han}
\bibitem{lhctriplet3}
  T.~Han, B.~Mukhopadhyaya, Z.~Si and K.~Wang,
  %``Pair production of doubly-charged scalars: Neutrino mass constraints and signals at the LHC,''
  Phys.Rev.D{\bf 76},075013 (2007).
  %doi:10.1103/PhysRevD.76.075013
 % [arXiv:0706.0441 [hep-ph]].
  %%CITATION = doi:10.1103/PhysRevD.76.075013;%%
  %143 citations counted in INSPIRE as of 31 Oct 2016

   %\cite{Martin:2014nqa}
\bibitem{Martin:2014nqa}
  A. D. Martin and M. G. Ryskin,
  %``The photon PDF of the proton,''
  Eur.Phys.J.C {\bf 74}, 3040 (2014)
  %doi:10.1140/epjc/s10052-014-3040-y
 % [arXiv:1406.2118 [hep-ph]].
  %%CITATION = doi:10.1140/epjc/s10052-014-3040-y;%%
  %26 citations counted in INSPIRE as of 04 Nov 2016

%\cite{Manohar:2016nzj}
\bibitem{Manohar:2016nzj}
  A. Manohar, P. Nason, G. P. Salam and G. Zanderighi,
  %``How bright is the proton? A precise determination of the photon PDF,''
  arXiv:1607.04266 [hep-ph].
  %%CITATION = ARXIV:1607.04266;%%
  %9 citations counted in INSPIRE as of 04 Nov 2016




%\cite{Fichet:2015vvy}
\bibitem{diphoton}
  S.~Fichet, G.~von Gersdorff and C.~Royon,
  %``Scattering light by light at 750 GeV at the LHC,''
  Phys.Rev.D {\bf 93}, no. 7, 075031 (2016);\\
  %doi:10.1103/PhysRevD.93.075031
  %[arXiv:1512.05751 [hep-ph]].
  %%CITATION = doi:10.1103/PhysRevD.93.075031;%%
  %200 citations counted in INSPIRE as of 18 Dec 2016
%\cite{Csaki:2015vek}
%\bibitem{Csaki:2015vek}
  C.~Csaki, J.~Hubisz and J.~Terning,
  %``Minimal model of a diphoton resonance: Production without gluon couplings,''
  Phys.Rev.D {\bf 93}, no. 3, 035002 (2016);\\
  %doi:10.1103/PhysRevD.93.035002
  %[arXiv:1512.05776 [hep-ph]]
  %%CITATION = doi:10.1103/PhysRevD.93.035002;%%
  %184 citations counted in INSPIRE as of 18 Dec 2016
%\cite{Csaki:2016raa}
%\bibitem{Csaki:2016raa}
  C.~Csaki, J.~Hubisz, S.~Lombardo and J.~Terning,
  %``Gluon versus photon production of a 750 GeV diphoton resonance,''
  Phys.Rev.D {\bf 93}, no. 9, 095020 (2016); \\
  %doi:10.1103/PhysRevD.93.095020
 % [arXiv:1601.00638 [hep-ph]].
  %%CITATION = doi:10.1103/PhysRevD.93.095020;%%
  %94 citations counted in INSPIRE as of 18 Dec 2016
%\cite{Abel:2016pyc}
%\bibitem{Abel:2016pyc}
  S.~Abel and V.~V.~Khoze,
  %``Photo-production of a 750 GeV di-photon resonance mediated by Kaluza-Klein leptons in the loop,''
  JHEP {\bf 1605}, 063 (2016);\\
 % doi:10.1007/JHEP05(2016)063
%  [arXiv:1601.07167 [hep-ph]].
  %%CITATION = doi:10.1007/JHEP05(2016)063;%%
  %42 citations counted in INSPIRE as of 18 Dec 2016
%\cite{Fichet:2016pvq}
%\bibitem{Fichet:2016pvq}
  S.~Fichet, G.~von Gersdorff and C.~Royon,
  %``Measuring the Diphoton Coupling of a 750 GeV Resonance,''
  Phys.Rev.Lett.{\bf 116}, no. 23, 231801 (2016);\\
 % doi:10.1103/PhysRevLett.116.231801
%  [arXiv:1601.01712 [hep-ph]].
  %%CITATION = doi:10.1103/PhysRevLett.116.231801;%%
  %65 citations counted in INSPIRE as of 18 Dec 2016
%\cite{Barrie:2016ntq}
%\bibitem{Barrie:2016ntq}
  N.~D.~Barrie, A.~Kobakhidze, M.~Talia and L.~Wu,
  %``750 GeV Composite Axion as the LHC Diphoton Resonance,''
  Phys.Lett.B {\bf 755}, 343 (2016);\\
  %doi:10.1016/j.physletb.2016.02.010
  %[arXiv:1602.00475 [hep-ph]].
  %%CITATION = doi:10.1016/j.physletb.2016.02.010;%%
  %41 citations counted in INSPIRE as of 18 Dec 2016
%\cite{Barrie:2016ndh}
%\bibitem{Barrie:2016ndh}
  N.~D.~Barrie, A.~Kobakhidze, S.~Liang, M.~Talia and L.~Wu,
  %``Heavy Leptonium as the Origin of the 750 GeV Diphoton Excess,''
  arXiv:1604.02803 [hep-ph];\\
  %%CITATION = ARXIV:1604.02803;%%
  %12 citations counted in INSPIRE as of 18 Dec 2016
%\cite{Ghosh:2016lnu}
%\bibitem{Ghosh:2016lnu}
  K.~Ghosh, S.~Jana and S.~Nandi,
  %``Neutrino Mass Generation and 750 GeV Diphoton excess via photon-photon fusion at the Large Hadron Collider,''
  arXiv:1607.01910 [hep-ph];\\
  %%CITATION = ARXIV:1607.01910;%%
  %5 citations counted in INSPIRE as of 18 Dec 2016
%\cite{Agarwalla:2016rmw}
%\bibitem{Agarwalla:2016rmw}
  S.~K.~Agarwalla, K.~Ghosh and A.~Patra,
  %``LHC diphoton excess in a left-right symmetric model with minimal dark matter,''
  arXiv:1607.03878 [hep-ph].
  %%CITATION = ARXIV:1607.03878;%%
  %2 citations counted in INSPIRE as of 18 Dec 2016

\bibitem{seesaw}
 % T.~P.~Cheng and L.~F.~Li,
  %``Neutrino Masses, Mixings and Oscillations in SU(2) x U(1) Models of Electroweak Interactions,''
 % Phys.Rev.D {\bf 22}, 2860 (1980); \\
  %doi:10.1103/PhysRevD.22.2860
  %%CITATION = doi:10.1103/PhysRevD.22.2860;%%
  %750 citations counted in INSPIRE as of 17 Dec 2016
%\cite{Magg:1980ut}
%\bibitem{Magg:1980ut}
  M.~Magg and C.~Wetterich,
  %``Neutrino Mass Problem and Gauge Hierarchy,''
  Phys.Lett.{\bf 94B}, 61 (1980); \\
  %doi:10.1016/0370-2693(80)90825-4
  %%CITATION = doi:10.1016/0370-2693(80)90825-4;%%
  %676 citations counted in INSPIRE as of 17 Dec 2016
%\cite{Schechter:1980gr}
%\bibitem{Schechter:1980gr}
  J.~Schechter and J.~W.~F.~Valle,
  %``Neutrino Masses in SU(2) x U(1) Theories,''
  Phys.Rev.D {\bf 22}, 2227 (1980); \\
  %doi:10.1103/PhysRevD.22.2227
  %%CITATION = doi:10.1103/PhysRevD.22.2227;%%
  %1996 citations counted in INSPIRE as of 17 Dec 2016
  %\cite{Lazarides:1980nt}
%\bibitem{Lazarides:1980nt}
  G.~Lazarides, Q.~Shafi and C.~Wetterich,
  %``Proton Lifetime and Fermion Masses in an SO(10) Model,''
  Nucl. Phys. B {\bf 181}, 287 (1981);\\
  %doi:10.1016/0550-3213(81)90354-0
  %%CITATION = doi:10.1016/0550-3213(81)90354-0;%%
  %1053 citations counted in INSPIRE as of 17 Dec 2016
%\cite{Mohapatra:1980yp}
%\bibitem{Mohapatra:1980yp}
  R.~N.~Mohapatra and G.~Senjanovic,
  %``Neutrino Masses and Mixings in Gauge Models with Spontaneous Parity Violation,''
  Phys.Rev.D {\bf 23}, 165 (1981).
  %doi:10.1103/PhysRevD.23.165
  %%CITATION = doi:10.1103/PhysRevD.23.165;%%
  %2040 citations counted in INSPIRE as of 17 Dec 2016


%\cite{Pati:1974yy}
\bibitem{lrsm}
  J. C. Pati and A. Salam,
  %``Lepton Number as the Fourth Color,''
  Phys.Rev.D {\bf 10}, 275(1974);\\
  %Erratum: [Phys.Rev.D {\bf 11},703(1975)].
  %doi:10.1103/PhysRevD.10.275,10.1103/PhysRevD.11.703.2
  %%CITATION = doi:10.1103/PhysRevD.10.275, 10.1103/PhysRevD.11.703.2;%%
  %4091 citations counted in INSPIRE as of 29 Oct 2016
%\cite{Mohapatra:1974hk}
%\bibitem{Mohapatra:1974hk}
  R.~N.~Mohapatra and J.~C.~Pati,
  %``Left-Right Gauge Symmetry and an Isoconjugate Model of CP Violation,''
  Phys.Rev.D {\bf 11}, 566 (1975);\\
  %doi:10.1103/PhysRevD.11.566
  %%CITATION = doi:10.1103/PhysRevD.11.566;%%
  %1834 citations counted in INSPIRE as of 29 Oct 2016
  %\cite{Senjanovic:1975rk}
%\bibitem{Senjanovic:1975rk}
  G.~Senjanovic and R.~N.~Mohapatra,
  %``Exact Left-Right Symmetry and Spontaneous Violation of Parity,''
  Phys.Rev.D{\bf 12}, 1502 (1975).
  %doi:10.1103/PhysRevD.12.1502
  %%CITATION = doi:10.1103/PhysRevD.12.1502;%%
  %1749 citations counted in INSPIRE as of 29 Oct 2016



%\cite{Kuchimanchi:1993jg}
\bibitem{susylrsm}
  R.~Kuchimanchi and R.~N.~Mohapatra,
  %``No parity violation without R-parity violation,''
  Phys.Rev.D {\bf 48}, 4352 (1993); \\
  %doi:10.1103/PhysRevD.48.4352
  %[hep-ph/9306290].
  %%CITATION = doi:10.1103/PhysRevD.48.4352;%%
  %159 citations counted in INSPIRE as of 24 Dec 2016
%\cite{Babu:2008ep}
%\bibitem{Babu:2008ep}
  K.~S.~Babu and R.~N.~Mohapatra,
  %``Minimal Supersymmetric Left-Right Model,''
  Phys.Lett.B {\bf 668}, 404 (2008);\\
%  doi:10.1016/j.physletb.2008.09.018
 % [arXiv:0807.0481 [hep-ph]].
  %%CITATION = doi:10.1016/j.physletb.2008.09.018;%%
  %44 citations counted in INSPIRE as of 24 Dec 2016
  %\cite{Babu:2014vba}
%\bibitem{Babu:2014vba}
  K.~S.~Babu and A.~Patra,
  %``Higgs Boson Spectra in Supersymmetric Left-Right Models,''
  Phys.\ Rev.\ D {\bf 93}, no. 5, 055030 (2016);
%  doi:10.1103/PhysRevD.93.055030
%  [arXiv:1412.8714 [hep-ph]];\\
  %%CITATION = doi:10.1103/PhysRevD.93.055030;%%
  %10 citations counted in INSPIRE as of 28 Dec 2016
  %\cite{Basso:2015pka}
%\bibitem{Basso:2015pka}
  L.~Basso, B.~Fuks, M.~E.~Krauss and W.~Porod,
  %``Doubly-charged Higgs and vacuum stability in left-right supersymmetry,''
  JHEP {\bf 1507}, 147 (2015).
%  doi:10.1007/JHEP07(2015)147
%  [arXiv:1503.08211 [hep-ph]].
  %%CITATION = doi:10.1007/JHEP07(2015)147;%%
  %7 citations counted in INSPIRE as of 28 Dec 2016

%\cite{Rizzo:1981xx}
%\bibitem{Rizzo:1981xx}
 % T.~G.~Rizzo,
  %``Doubly Charged Higgs Bosons and Lepton Number Violating Processes,''
  %Phys.Rev.D{\bf 25}, 1355 (1982)
 % Addendum: [Phys.Rev.D {\bf 27}, 657 (1983)].
  %doi:10.1103/PhysRevD.27.657, 10.1103/PhysRevD.25.1355
  %%CITATION = doi:10.1103/PhysRevD.27.657, 10.1103/PhysRevD.25.1355;%%
  %97 citations counted in INSPIRE as of 29 Oct 2016
%\cite{Cheng:1980qt}


%\cite{Zee:1985id}
\bibitem{Zee}
  A.~Zee,
  %``Quantum Numbers of Majorana Neutrino Masses,''
  Nucl.Phys.B {\bf 264}, 99 (1986); \\
  %doi:10.1016/0550-3213(86)90475-X
  %%CITATION = doi:10.1016/0550-3213(86)90475-X;%%
  %315 citations counted in INSPIRE as of 29 Oct 2016
%\cite{Babu:1988ki}
%\bibitem{Babu:1988ki}
  K. S. Babu, Phys.Lett.B {\bf 203}, 132 (1988).
  %doi:10.1016/0370-2693(88)91584-5
   %``Model of 'Calculable' Majorana Neutrino Masses,''
   %%CITATION = doi:10.1016/0370-2693(88)91584-5;%%
  %508 citations counted in INSPIRE as of 29 Oct 2016
%\cite{CiezaMontalvo:2006zt}

\bibitem{modela}
  N.~Arkani-Hamed, A.~G.~Cohen, E.~Katz, A.~E.~Nelson, T.~Gregoire and J.~G.~Wacker,
  %``The Minimal moose for a little Higgs,''
  JHEP {\bf 0208}, 021 (2002).
  %doi:10.1088/1126-6708/2002/08/021
 % [hep-ph/0206020].
  %%CITATION = doi:10.1088/1126-6708/2002/08/021;%%
  %588 citations counted in INSPIRE as of 29 Oct 2016

  %\cite{Georgi:1985nv}
\bibitem{modelb}
  H.~Georgi and M.~Machacek,
  %``Doubly Charged Higgs Bosons,''
  Nucl.Phys.B {\bf 262}, 463 (1985);\\
  %doi:10.1016/0550-3213(85)90325-6
  %%CITATION = doi:10.1016/0550-3213(85)90325-6;%%
  %266 citations counted in INSPIRE as of 29 Oct 2016
  %\cite{Gunion:1989ci}
%\bibitem{c}
  J.~F.~Gunion, R.~Vega and J.~Wudka,
  %``Higgs triplets in the standard model,''
  Phys.Rev.D {\bf 42}, 1673 (1990).
  %doi:10.1103/PhysRevD.42.1673
  %%CITATION = doi:10.1103/PhysRevD.42.1673;%%
  %239 citations counted in INSPIRE as of 29 Oct 2016
  %\cite{ArkaniHamed:2002qx}

\bibitem{modelc}
  K.~S.~Babu, S.~Nandi and Z.~Tavartkiladze,
  %``New Mechanism for Neutrino Mass Generation and Triply Charged Higgs Bosons at the LHC,''
  Phys. Rev.D {\bf 80}, 071702 (2009);\\
  %doi:10.1103/PhysRevD.80.071702
  %[arXiv:0905.2710 [hep-ph]].
  %%CITATION = doi:10.1103/PhysRevD.80.071702;%%
  %77 citations counted in INSPIRE as of 29 Oct 2016
  %\cite{Bonnet:2009ej}
%\bibitem{Bonnet:2009ej}
  F.~Bonnet, D.~Hernandez, T.~Ota and W.~Winter,
  %``Neutrino masses from higher than d=5 effective operators,''
  JHEP {\bf 0910}, 076 (2009);\\
  %doi:10.1088/1126-6708/2009/10/076
 % [arXiv:0907.3143 [hep-ph]].
  %%CITATION = doi:10.1088/1126-6708/2009/10/076;%%
  %83 citations counted in INSPIRE as of 25 Dec 2016
%\cite{Kumericki:2012bh}
%\bibitem{Kumericki:2012bh}
  K.~Kumericki, I.~Picek and B.~Radovcic,
  %``TeV-scale Seesaw with Quintuplet Fermions,''
  Phys.Rev.D {\bf 86}, 013006 (2012).
  %doi:10.1103/PhysRevD.86.013006
 % [arXiv:1204.6599 [hep-ph]].
  %%CITATION = doi:10.1103/PhysRevD.86.013006;%%
  %63 citations counted in INSPIRE as of 25 Dec 2016


\bibitem{lhctriplet1}
  A.~G.~Akeroyd and M.~Aoki,
  %``Single and pair production of doubly charged Higgs bosons at hadron colliders,''
  Phys.Rev.D {\bf 72}, 035011 (2005);\\
  %[hep-ph/0506176].
  %%CITATION = doi:10.1103/PhysRevD.72.035011;%%
  %167 citations counted in INSPIRE as of 29 Oct 2016
  %\cite{Hektor:2007uu}
%\bibitem{lhc3}
  P.~Fileviez Perez, T.~Han, G.~Y.~Huang, T.~Li and K.~Wang,
  %``Testing a Neutrino Mass Generation Mechanism at the LHC,''
  Phys.Rev. D {\bf 78}, 071301 (2008); \\
  %doi:10.1103/PhysRevD.78.071301
%  [arXiv:0803.3450 [hep-ph]].
  %%CITATION = doi:10.1103/PhysRevD.78.071301;%%
  %61 citations counted in INSPIRE as of 29 Oct 2016
   % \cite{}
%\bibitem{lhc4}
  W.~Chao, Z.~G.~Si, Z.~z.~Xing and S.~Zhou,
  %``Correlative signatures of heavy Majorana neutrinos and doubly-charged Higgs bosons at the Large Hadron Collider,''
  Phys.Lett.B {\bf 666}, 451 (2008); \\
  %doi:10.1016/j.physletb.2008.08.003
 % [arXiv:0804.1265 [hep-ph]].
  %%CITATION = doi:10.1016/j.physletb.2008.08.003;%%
  %48 citations counted in INSPIRE as of 29 Oct 2016
  %\cite{Nebot:2007bc}
%\cite{Chun:2003ej}
%\bibitem{lhc14}
  E.~J.~Chun, K.~Y.~Lee and S.~C.~Park,
  %``Testing Higgs triplet model and neutrino mass patterns,''
  Phys.Lett.B {\bf 566}, 142 (2003);\\
  %doi:10.1016/S0370-2693(03)00770-6
  %[hep-ph/0304069].
  %%CITATION = doi:10.1016/S0370-2693(03)00770-6;%%
  %155 citations counted in INSPIRE as of 10 Dec 2016
%\cite{Perez:2008ha}
%\bibitem{lhc15}
  P.~Fileviez Perez, T.~Han, G.~y.~Huang, T.~Li and K.~Wang,
  %``Neutrino Masses and the CERN LHC: Testing Type II Seesaw,''
  Phys.Rev.D {\bf 78}, 015018 (2008); \\
  %doi:10.1103/PhysRevD.78.015018
 % [arXiv:0805.3536 [hep-ph]].
  %%CITATION = doi:10.1103/PhysRevD.78.015018;%%
  %230 citations counted in INSPIRE as of 10 Dec 2016
  %\cite{Kanemura:2013vxa}
  %\cite{Mitra:2016wpr}
%\bibitem{lhc19}
  M.~Mitra, S.~Niyogi and M.~Spannowsky,
  %``Type-II Seesaw and Multilepton Signatures at Hadron Colliders,''
 arXiv:1611.09594 [hep-ph];\\
  %%CITATION = ARXIV:1611.09594;%%
  %\bibitem{lhc2}
  S.~Kanemura, K.~Yagyu and H.~Yokoya,
  %``First constraint on the mass of doubly-charged Higgs bosons in the same-sign diboson decay scenario at the LHC,''
  Phys.Lett.B {\bf 726}, 316 (2013); \\
  %doi:10.1016/j.physletb.2013.08.054
 % [arXiv:1305.2383 [hep-ph]].
  %%CITATION = doi:10.1016/j.physletb.2013.08.054;%%
  %54 citations counted in INSPIRE as of 10 Dec 2016
%\cite{Muhlleitner:2003me}
%\bibitem{lhc2}
  M.~Muhlleitner and M.~Spira,
  %``A Note on doubly charged Higgs pair production at hadron colliders,''
  Phys.Rev. D {\bf 68}, 117701 (2003).
  %doi:10.1103/PhysRevD.68.117701
  %[hep-ph/0305288].
  %%CITATION = doi:10.1103/PhysRevD.68.117701;%%
  %122 citations counted in INSPIRE as of 29 Oct 2016
  %\cite{Akeroyd:2005gt}


%\cite{Melfo:2011nx} : electroweak precision test
\bibitem{lhctriplet2}
  A.~Melfo, M.~Nemevsek, F.~Nesti, G.~Senjanovic and Y.~Zhang,
  %``Type II Seesaw at LHC: The Roadmap,''
  Phys.Rev.D {\bf 85}, 055018 (2012).
  %doi:10.1103/PhysRevD.85.055018
  %[arXiv:1108.4416 [hep-ph]].
  %%CITATION = doi:10.1103/PhysRevD.85.055018;%%
  %99 citations counted in INSPIRE as of 31 Oct 2016



\bibitem{lhcrad}
  M.~Nebot, J.~F.~Oliver, D.~Palao and A.~Santamaria,
  %``Prospects for the Zee-Babu Model at the CERN LHC and low energy experiments,''
  Phys.Rev.D {\bf 77}, 093013 (2008); \\
  %doi:10.1103/PhysRevD.77.093013
 % [arXiv:0711.0483 [hep-ph]].
  %%CITATION = doi:10.1103/PhysRevD.77.093013;%%
  %84 citations counted in INSPIRE as of 29 Oct 2016
%\cite{Gunion:1989in}
%\bibitem{lhc81}
  M.~Nebot, J.~F.~Oliver, D.~Palao and A.~Santamaria,
  %``Prospects for the Zee-Babu Model at the CERN LHC and low energy experiments,''
  Phys.Rev.D {\bf 77}, 093013 (2008).

 \bibitem{lhclrsm}
  J.~F.~Gunion, J.~Grifols, A.~Mendez, B.~Kayser and F.~I.~Olness,
  %``Higgs Bosons in Left-Right Symmetric Models,''
  Phys.Rev.D {\bf 40}, 1546 (1989); \\
  %doi:10.1103/PhysRevD.40.1546
  %%CITATION = doi:10.1103/PhysRevD.40.1546;%%
  %235 citations counted in INSPIRE as of 29 Oct 2016
  %\cite{Deshpande:1990ip}
  %\bibitem{lhc7}
  K.~Huitu, J.~Maalampi, A.~Pietila and M.~Raidal,
  %``Doubly charged Higgs at LHC,''
  Nucl.Phys.B {\bf 487}, 27 (1997);\\
  %doi:10.1016/S0550-3213(97)87466-4
  %[hep-ph/9606311].
  %%CITATION = doi:10.1016/S0550-3213(97)87466-4;%%
  %189 citations counted in INSPIRE as of 29 Oct 2016
  %\cite{Babu:2009aq}
%\cite{Bambhaniya:2015wna}
%\bibitem{lhc20}
  G.~Bambhaniya, J.~Chakrabortty, J.~Gluza, T.~Jelinski and R.~Szafron,
  %``Search for doubly charged Higgs bosons through vector boson fusion at the LHC and beyond,''
  Phys.Rev.D{\bf 92},no. 1,015016 (2015); \\
  %doi:10.1103/PhysRevD.92.015016
 % [arXiv:1504.03999 [hep-ph]].
  %%CITATION = doi:10.1103/PhysRevD.92.015016;%%
  %14 citations counted in INSPIRE as of 31 Oct 2016
%\cite{Dutta:2014dba}
%\bibitem{lhc21}
  B.~Dutta, R.~Eusebi, Y.~Gao, T.~Ghosh and T.~Kamon,
  %``Exploring the doubly charged Higgs boson of the left-right symmetric model using vector boson fusionlike events at the LHC,''
  Phys.Rev.D {\bf 90}, 055015 (2014); \\
  %doi:10.1103/PhysRevD.90.055015
 % [arXiv:1404.0685 [hep-ph]].
  %%CITATION = doi:10.1103/PhysRevD.90.055015;%%
  %14 citations counted in INSPIRE as of 31 Oct 2016
  %\cite{Babu:2013ega}
%\bibitem{lhc}
  K.~S.~Babu, A.~Patra and S.~K.~Rai,
  %``New Signals for Doubly-Charged Scalars and Fermions at the Large Hadron Collider,''
  Phys.Rev.D {\bf 88}, 055006 (2013).
%  doi:10.1103/PhysRevD.88.055006
 % [arXiv:1306.2066 [hep-ph]].
  %%CITATION = doi:10.1103/PhysRevD.88.055006;%%
  %12 citations counted in INSPIRE as of 24 Dec 2016

\bibitem{lhclittleh}
  A.~Hektor, M.~Kadastik, M.~Muntel, M.~Raidal and L.~Rebane,
  %``Testing neutrino masses in little Higgs models via discovery of doubly charged Higgs at LHC,''
  Nucl.Phys.B {\bf 787}, 198 (2007).
  %doi:10.1016/j.nuclphysb.2007.07.014
  %[arXiv:0705.1495 [hep-ph]].
  %%CITATION = doi:10.1016/j.nuclphysb.2007.07.014;%%
  %86 citations counted in INSPIRE as of 29 Oct 2016
%\cite{Perez:2008zc}

\bibitem{lhcother}
  J.~E.~Cieza Montalvo, N.~V.~Cortez, J.~Sa Borges and M.~D.~Tonasse,
  %``Searching for doubly charged Higgs bosons at the LHC in a 3-3-1 model,''
  Nucl.Phys.B {\bf 756}, 1(2006); \\
  %Erratum: [Nucl.Phys.B {\bf 796}, 422(2008)]
  %doi:10.1016/j.nuclphysb.2006.08.013, 10.1016/j.nuclphysb.2008.01.003
 % [hep-ph/0606243].
  %%CITATION = doi:10.1016/j.nuclphysb.2006.08.013, 10.1016/j.nuclphysb.2008.01.003;%%
  %35 citations counted in INSPIRE as of 29 Oct 2016
%\bibitem{lhcquadruplet}
  S.~Bhattacharya, S.~Jana and S.~Nandi,
  %``Neutrino Masses and Scalar Singlet Dark Matter,''
  arXiv:1609.03274 [hep-ph]; \\
  %%CITATION = ARXIV:1609.03274;%%
%\cite{Gunion:1996pq}
%\bibitem{lhcgen2}
  J.~F.~Gunion, C.~Loomis and K.~T.~Pitts,
  %``Searching for doubly charged Higgs bosons at future colliders,''
  eConf C {\bf 960625}, LTH096 (1996);\\
  %[hep-ph/9610237].
  %%CITATION = HEP-PH/9610237;%%
  %99 citations counted in INSPIRE as of 12 Dec 2016
%\cite{Rentala:2011mr}
%\bibitem{lhcgen3}
  V.~Rentala, W.~Shepherd and S.~Su,
  %``A Simplified Model Approach to Same-sign Dilepton Resonances,''
  Phys.Rev.D {\bf 84}, 035004 (2011).
  %doi:10.1103/PhysRevD.84.035004
  %[arXiv:1105.1379 [hep-ph]].
  %%CITATION = doi:10.1103/PhysRevD.84.035004;%%
  %26 citations counted in INSPIRE as of 12 Dec 2016




%\cite{CMS:2016cpz}
\bibitem{CMS:2016cpz}
  CMS Collaboration [CMS Collaboration],
  %``Search for a doubly-charged Higgs boson with $\sqrt{s}=8~\mathrm{TeV}$ $pp$ collisions at the CMS experiment,''
  CMS-PAS-HIG-14-039.
  %%CITATION = CMS-PAS-HIG-14-039;%%
  %5 citations counted in INSPIRE as of 01 Nov 2016




  %\cite{Belyaev:2012qa} :calchep
\bibitem{Belyaev:2012qa}
  A.~Belyaev, N.~D.~Christensen and A.~Pukhov,
  %``CalcHEP 3.4 for collider physics within and beyond the Standard Model,''
  Comput.Phys.Commun.{\bf 184}, 1729 (2013)
  %doi:10.1016/j.cpc.2013.01.014
  %[arXiv:1207.6082 [hep-ph]].
  %%CITATION = doi:10.1016/j.cpc.2013.01.014;%%
  %369 citations counted in INSPIRE as of 31 Oct 2016






\end{thebibliography}
\end{document}